\documentclass{article}

\usepackage{arxiv}

\usepackage[utf8]{inputenc} % allow utf-8 input
\usepackage[T1]{fontenc}    % use 8-bit T1 fonts
\usepackage{hyperref}       % hyperlinks
\usepackage{url}            % simple URL typesetting
\usepackage{booktabs}       % professional-quality tables
\usepackage{amsfonts}       % blackboard math symbols
\usepackage{nicefrac}       % compact symbols for 1/2, etc.
\usepackage{microtype}      % microtypography
\usepackage{lipsum}
\usepackage{graphicx}
\graphicspath{ {./images/} }
\usepackage{amsmath}
\usepackage{amsmath,amssymb,amsfonts}
\usepackage{bm}
\usepackage[linesnumbered,ruled]{algorithm2e}
\usepackage{color}
\usepackage{multirow}
\usepackage{makecell}
\usepackage{booktabs}
\usepackage[T1]{fontenc}
\usepackage{lineno}
\usepackage{soul}
\usepackage{pifont}
\usepackage{multirow}

\title{Enhanced 3D Gravity Inversion Using ResU-Net with Density Logging Constraints: A Dual-Phase Training Approach}

\author{
 Siyuan Dong \\
  School of Information and Communications Engineering\\
  Xi'an Jiaotong University\\
  Xi'an, Shanxi, China \\
  \texttt{sydong24@stu.xjtu.edu.cn} \\
   \And
 Jinghuai Gao \\
  School of Information and Communications Engineering\\
  Xi'an Jiaotong University\\
  Xi'an, Shanxi, China \\
  \texttt{jhgao@mail.xjtu.edu.cn} \\
  \And
 Shuai Zhou \\
  College of Geoexploration Science and Technology\\
  Jilin University\\
  Changchun, Jilin, China \\
  \texttt{zhoushuai@jlu.edu.cn} \\
   \And
Baohai Wu \\
School of Information and Communications Engineering\\
Xi'an Jiaotong University\\
Xi'an, Shanxi, China \\
\texttt{baohaiwu@163.com} \\
\And
Hongfa Jia \\
College of Geoexploration Science and Technology\\
Jilin University\\
Changchun, Jilin, China \\
\texttt{jiahf24@mails.jlu.edu.cn} \\
}

\begin{document}
\maketitle
\begin{abstract}
Gravity exploration has become an important geophysical method due to its low cost and high efficiency. With the rise of artificial intelligence, data-driven gravity inversion methods based on deep learning (DL) possess physical property recovery capabilities that conventional regularization methods lack. However, existing DL methods suffer from insufficient prior information constraints, which leads to inversion models with large data fitting errors and unreliable results. Moreover, the inversion results lack constraints and matching from other exploration methods, leading to results that may contradict known geological conditions.
In this study, we propose a novel approach that integrates prior density well logging information to address the above issues. First, we introduce a depth weighting function to the neural network (NN) and train it in the weighted density parameter domain. The NN, under the constraint of the weighted forward operator, demonstrates improved inversion performance, with the resulting inversion model exhibiting smaller data fitting errors. Next, we divide the entire network training into two phases: first training a large pre-trained network Net-I, and then using the density logging information as the constraint to get the optimized fine-tuning network Net-II. Through testing and comparison in synthetic models and Bishop Model, the inversion quality of our method has significantly improved compared to the unconstrained data-driven DL inversion method. Additionally, we also conduct a comparison and discussion between our method and both the conventional focusing inversion (FI) method and its well logging constrained variant. Finally, we apply this method to the measured data from the San Nicolas mining area in Mexico, comparing and analyzing it with two recent gravity inversion methods based on DL.
\end{abstract}

% keywords can be removed
%\keywords{First keyword \and Second keyword \and More}

\section{Introduction}
\label{sec:intro}
Modern gravity exploration, with its advantages of low cost, high efficiency, and ease of implementation, is highly applicable in fields such as strategic resource exploration \cite{yang2019application}, geothermal reservoir estimation \cite{pearson2018gravity}, and geological structure studies \cite{geng2019gravity}.

As one of the core issues in gravity exploration, gravity inversion has seen significant development through the introduction of various constraint conditions. The core concept of sparsity constraints is to seek a solution with the minimum number of non-zero elements, thereby obtaining compact inversion results. \cite{last1983compact} introduced a volume minimization constraint based on the L0 norm in the loss function to enforce minimal model volume for achieving sparsity, thereby producing focused and compact inversion results. \cite{sun2014adaptive} introduced an adaptive Lp-norm scheme, which dynamically adjusts the norm exponent $p$ in different regions of the model, enabling the recovery of both blocky and smooth features within a single inversion. In contrast, focusing constraints require the gradient of the model parameters to be sparse, thereby maintaining sharp geological boundaries. \cite{portniaguine1999focusing} proposed a minimal gradient support constraint, whose core principle is to minimize regions with non-zero model parameter gradients. This constraint restricts gradients to exist only within extremely narrow zones of abrupt variations, ensuring sharp boundaries and maximally flat interiors in the inversion model. Explicit prior information serves as another powerful category of constraints to reduce non-uniqueness. In the inversion of three-dimensional (3-D) crustal structure beneath central Taiwan, \cite{jian2004three} leveraged known seismic velocity data to construct the initial density model of the crust.  \cite{krahenbuhl2006inversion} utilized structural information of the salt dome's top and known density data, employing a binary inversion approach to highlight subsurface property contrasts, which effectively reconstructed the structure beneath top of salt. \cite{silva20093d} integrated spatial geometric configurations and density information, employing an adaptive algorithm to iteratively refine coarse subsurface meshes, thereby achieving finely resolved and compact inversion results. \cite{zhdanov2017adaptive} proposed an adaptive multinary inversion approach that utilizes prior subsurface density information to discretizes the continuous density distribution, thereby forcing the inverted model to converge toward predefined discrete density values. Moreover, incorporating depth weighting function as constraint serves as a critical strategy for enhancing depth resolution of gravity inversion \cite{li19983}.

Low depth resolution is an inherent limitation of gravity inversion, mainly due to the skin effect caused by the weight decay of the sensitivity matrix with increasing distance. The depth weighting method is specifically developed to address this issue, primarily through two approaches. First, modifying the loss function by increasing penalty strength for shallow structures while reducing penalties for deep regions to balance physical property distribution. Second, directly optimizing the sensitivity matrix to enhance the weight for deeper zones. In the first category, \cite{li19963} proposed a depth-varying weighting function whose parameters are tuned to approximate the decay rate of the kernel function, thereby improving inversion depth resolution. Building on this, \cite{commer2011three} developed a spatial weighting function that adjusts penalty strengths at different depths based on approximate burial depths and vertical distribution patterns of the model. However, these methods have a certain degree of subjectivity in parameter selection, where inappropriate choices directly compromise inversion quality. \cite{cella2012inversion} argued that the methods based on the theory of homogeneous fields (i.e., Euler deconvolution) and depth weighting function have a significant link, and can be used to estimate the construction index N for more objective parameter selection. The second category fundamentally involves computing the L2 norm of weights from each subsurface point to all observation points, thereby rebalancing the sensitivity matrix's weight at different depth to achieve effects comparable to the first category \cite{portniaguine1999focusing, li2000joint}. These methods rely solely on the sensitivity matrix values themselves, minimizing subjective interference, though at the cost of increased computational overhead. In summary, the core rationale of depth weighting lies in compensating for the low sensitivity of deep model parameters, representing one of the most critical approaches for enhancing depth resolution in gravity inversion.

While depth weighting effectively counteracts the natural resolution decay with depth, it represents a generic prior that cannot uniquely determine absolute physical property values within the model. Well logging data, serving as a complementary information source, provides more specific subsurface knowledge and further constrains the solution space \cite{lines1988cooperative, sun2008density, vasiljevic2019simple}. Consequently, depth weighting and well logging constraints act synergistically, the former establishes a stable operator with enhanced depth weighting, while the latter anchors the model to known physical properties at specific locations. Together, they produce a geologically more plausible and unique inversion result. \cite{chasseriau20033d} introduced density well logging data into 3-D gravity inversion to better determine the geometry and location of the inversion target. \cite{liu2022structure} used the density points provided by density well logging as constraint conditions to improve the depth resolution. \cite{liu2025petrophysical} used prior knowledge obtained from well logging as the initial model to constrain the gravity inversion. However, well logging constraints are inherently a form of sparse local control, while depth weighting constitutes a generic averaging compensation. Both strategies lack the capability to reconstruct structural information in deep regions. Consequently, conventional inversion relying on the aforementioned regularization methods exhibits an inherent upper limit in resolution enhancement. To break through this constraint and obtain sharper inversion models, the introduction of a fundamentally new approach is required.

Nowadays, DL has become an important and widely applied tool in the field of geophysics \cite{gao2019optimized, meng2022seismic, chen2023nonstationary, chenhongling2024, chen2025unsupervised, zhou2025seismic}. In gravity inversion, this data-driven training process can achieve depth resolution that conventional regularization methods lack, offering advantages in physical property recovery \cite{he2021recovering, zhang2021joint, zhang2022decnet}. \cite{huang2021deep} implemented 3-D gravity sparse inversion based on U-Net. \cite{lv2023fast} proposed a multitask UNet3+ to realize anomaly bodies localization and density contrasts reconstruction simultaneously. However, the aforementioned unconstrained DL inversion disregards data fitting by merely minimizing model errors. Lacking explicit physical constraints that guide how density at different depths influences gravity anomalies, the NN cannot discern whether residuals in shallow or deep zones of the model require higher priority during optimization. This leads to significant data misfit and undermines the physical plausibility of the results. \cite{jiao20233} argued that while NN exhibit superior fitting capabilities, learning the complex relationships between arbitrary subsurface models and their responses from limited training datasets remains highly challenging, requiring the incorporation of additional constraints to reduce the solution space. Consequently, integrating prior knowledge and physical constraints into NN has emerged as a critical strategy in recent years to mitigate solution non-uniqueness and enhance inversion performance. \cite{colombo2021physics} proposed a physics-driven coupled DL inversion framework. This approach modularizes and parallelizes regularized inversion and DL inversion based on joint inversion principles, where the two methods compete to obtain optimal solutions while iteratively refining each other's models through successive iterations. \cite{dong20233} introduced forward fitting as a constraint term into the NN to reduce data misfit in inversion results, followed by fine-tuning the network with increased forward-fitting weights to enhance alignment with observed data. Meanwhile, \cite{zhou2024magnetic} augmented boundary detection with depth cues, which not only enriches boundary features but also improves depth resolution.

Currently, DL methods in gravity inversion generally lack integration with physical constraints, limiting their applicability to complex geological scenarios. Meanwhile, multi-physics constrained inversion has emerged as a prominent research focus and developmental trend, demonstrating theoretical and practical effectiveness in mitigating solution non-uniqueness and enhancing inversion quality. To address these challenges, we propose a DL gravity inversion method that integrates prior density logging information to obtain results that are more consistent with the known geological conditions. The method employs ResU-Net architecture \cite{xiao2018weighted}, which offers both training stability and computational efficiency for this inversion task. The specific details are as follows: 

(1) By incorporating depth weighting into the NN, a pre-trained network (Net-I) is obtained through training on datasets in the weighted density parameter domain. Depth weighting provides the NN with a physically-guided sensitivity operator that assigns more balanced weights to deeper zones. Compared to unconstrained DL method, this approach yields inversion results with improved model consistency and data consistency, reflected in average increases of 7.4$\%$ and 6.8$\%$ in the relevant metrics, respectively.

(2) Building upon Net-I, fine-tuning with sparse prior well logging constraints produces Net-II. The spatial and physical property information serves to anchor and calibrate the inversion results of Net-I, ensuring full alignment with the known well logging data. Relative to Net-I, this further enhances both model consistency and data consistency in the inversion results, with average improvements of 3.0$\%$ and 2.1$\%$ in the corresponding metrics.

Through comparisons with synthetic models and Bishop Model, we verify that the proposed method exhibits strong generalization capabilities and superior depth resolution. Subsequently, in the comparison with the FI method and its well logging constrained variant, we discuss the tasks for which our method is more suitable. Finally, we apply this method to the field data from the San Nicolas mining area in central Mexico and compare it with two of the latest gravity inversion methods based on DL, proving that our method's inversion results are more consistent with the known geological conditions and mainstream interpretations.

\section{Methodology} \label{sec:method}
\subsection*{Fast Forward Modeling Based on Geometric Trellis}
\label{subsec:fast_forward}
The purpose of gravity forward modeling is to calculate the response caused by the underground density distribution at the surface \cite{chen2019fast, hogue2020tutorial}. The expression for the gravity anomaly $ \Delta g$ of a single rectangular model cell $\sigma (\xi_i, \eta_j, \zeta_k)$ relative to the observation point $P(x, y)$ is as follows \cite{li1998three}:
\begin{linenomath*}
	\begin{align}\label{eq1}
		\Delta g & = -G\rho\sum^2_{i=1}\sum^2_{j=1}\sum^2_{k=1}(-1)^{i+j+k}\cdot
		\notag
		\\ & \left[x_i\ln(y_i+r_{ijk})+y_i\ln(x_i+r_{ijk})-z_k\arctan(\frac{x_i y_j}{z_k r_{ijk}})\right]
	\end{align}
\end{linenomath*}
where $G$ is the gravitational constant, which is typically set to $G = 6.67\times10^{-11}\rm{m^3/(kg\cdot{s^2}})$, $\rho$ is the density of the rectangular model unit, $x_i = (x - \xi_i)$ , $y_j = (y - \eta_j)$ , $z_k = (z - \zeta_k)$ , and $r_{ijk} = (x^2_i + y^2_j + z^2_k)^{1/2}$. Equation \ref{eq1} can be simplified as follows: 
\begin{linenomath*}
	\begin{align}\label{eq2}
		\Delta g = \rho \cdot S(\xi_i, \eta_j, \zeta_k,x,y),
	\end{align}
\end{linenomath*}
where $S(\xi_i, \eta_j, \zeta_k,x,y)$ is also known as the geometric trellis \cite{yao2003high}, which represents the correspondence of any single rectangular model cell $\sigma(\xi_i, \eta_j, \zeta_k)$ at the grid point $P(x, y)$. The fast forward modeling code is available in the "Data Availability Statement" section.

\subsection*{Inversion Methodology Based on Depth Weighting Function}
\label{subsec:inv}
The linear gravity inversion problem, like that in Equation \ref{eq2}, can be simplified as follows:
\begin{linenomath*}
	\begin{align}\label{eq4}
		{\bf d} = {\bf A} {\bf m},
	\end{align}
\end{linenomath*}
where $ {\bf m} $ is the density model vector, $ {\bf d} $ is the anomaly data vector, and $ {\bf A} $ is the sensitivity matrix. Solving the inverse of a high-dimensional matrix $ {\bf A} $ is computationally expensive. However, the DL  can approximate the pseudo-inverse of matrix $ {\bf A}^{\dagger} $ through observation data ${\bf d}$ and NN parameters $\boldsymbol {\theta}$ to derive the learned inversion model ${\bf m}_l$ (an approximation of $ {\bf m} $), making it highly suitable for solving geophysical inversion problem. The specific expression is as follows:
\begin{linenomath*}
	\begin{align}\left\{\begin{aligned}\label{eq5}
			{\bf m} & = {\bf A}^{\dagger} {\bf d}
			\\ {\bf m}_l & = \rm{NNs} ({\bf d}, \boldsymbol {\theta}).
		\end{aligned}\right.\end{align}
\end{linenomath*}

Since the unconstrained DL method solely minimizes model regularization errors $\Vert {\bf m} - {\bf m}_l \Vert^2$ \cite{huang2021deep, he2021recovering, zhang2021deep, zhang2021joint, zhang2022decnet, lv2023fast}, it inevitably introduces significant data misfit in inversion results. Incorporating data constraint into the NNs can effectively mitigate this issue \cite{hu2024three}. Guided by this rationale, the loss function based on Dice function \cite{milletari2016v} in this study can be written as:
\begin{linenomath*}
	\begin{align}\label{eq6}
		\varPhi({\bf m}_l) = 
		1 - \left[\frac{{\bf m} \cdot {\bf m}_l}
		{\Vert {\bf m} \Vert^2 + \Vert {\bf m}_l \Vert^2} + 
		\frac{{\bf d} \cdot {\bf A}{\bf m}_l}
		{\Vert {\bf d} \Vert^2 + \Vert {\bf Am}_l \Vert^2}\right].
	\end{align}
\end{linenomath*}

\begin{figure}
	\noindent\includegraphics[width=\textwidth]{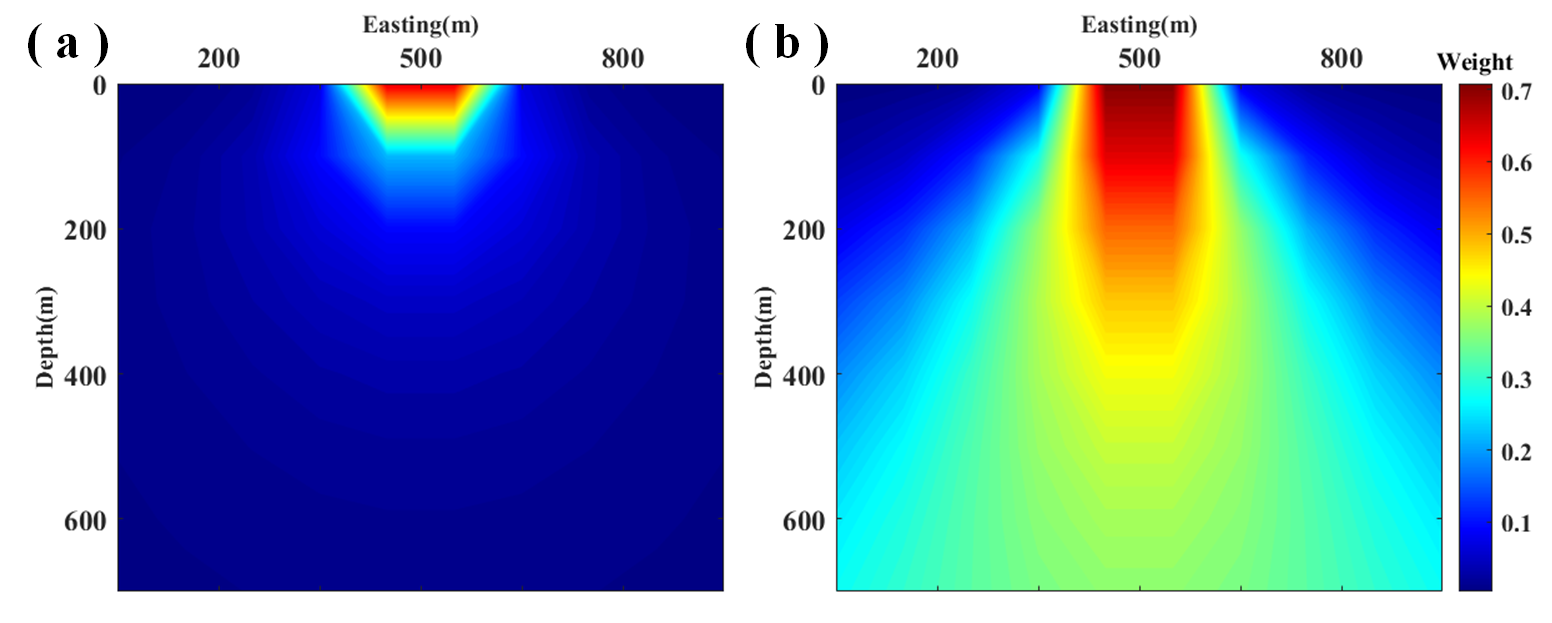}
	\caption{Visualization of the sensitivity matrix (the example grid $\Delta$ = 100m). During forward modeling, the weight distribution corresponding to a surface observation point at different depths is as follows: (a) ${\bf A}$ without weighting, (b)  ${\bf A}_w$ with depth weighting.}
	\label{fg2}
\end{figure}

However, the sensitivity matrix $ {\bf A} $ exhibits natural decay of weight with increasing depth. The specific weight distribution shows that the weights are larger at shallow depths and smaller at deeper depths (Figure \ref{fg2}a). Consequently, when the NN minimizes data misfit, the gradients for shallow zones exceed those for deep zones. This leads to inversion results that, while better fitting the observed anomalies, have negligible impact on the recovery of deep structures.

To better balance the contribution of different depths in the inversion and provide improved physical guidance to the NN, we introduce the depth weighting \cite{portniaguine20023}, which comprises components model weighting matrix ${\bf W}_m$, data weighting matrix ${\bf W}_d$ ,and the weighted space:
\begin{linenomath*}
	\begin{align}\left\{\begin{aligned}\label{eq7}
			{\bf W}_d & = {\rm diag}({\bf A} {\bf A}^T)^{1/2}\\
			{\bf W}_m & = {\rm diag}({\bf A}^T {\bf A})^{1/2}.
		\end{aligned}\right.
	\end{align}
\end{linenomath*}

\begin{linenomath*}
	\begin{align}\left\{\begin{aligned}\label{eq9}
			{\bf m}_w & = {\bf W}_m {\bf m} \\
			{\bf A}_w & = {\bf W}_d {\bf A} {\bf W}^{-1}_m. \\
			{\bf d}_w & = {\bf W}_d {\bf d}
		\end{aligned}\right.\end{align}
\end{linenomath*}

The matrix ${\bf A}$ is left-multiplied by matrix ${\bf W}_d$ to emphasize high-sensitivity observations on the data side, and then right-multiplied by matrix ${\bf W}^{-1}_m$ for column normalization, resulting in depth weighting sensitivity matrix ${\bf A}_w$. This process not only highlights informative data but also balance the depth weight distribution to avoid skin effect. The visualization of matrix ${\bf A}_w$ is shown in Figure \ref{fg2}b, where the weight of the deep part becomes higher. This indicates that during back-propagation, the NN will compute stronger gradients for deep structures, thereby enhancing their recovery in the inversion results. By combining Equation \ref{eq9}, we transform Equation \ref{eq6} into the weighted density parameter domain to solve for ${\bf m}_{wl}$ (an approximation of ${\bf m}_w$), resulting in a new loss function $\varPhi$ and its gradient with respect to ${\bf m}_{wl}$:

\begin{linenomath*}
	\begin{align}\label{eq10}
		\varPhi({\bf m}_{wl}) = 1 - 
		\left[\frac{{\bf m}_w \cdot {\bf m}_{wl}}
		{\Vert {\bf m}_w \Vert^2 + \Vert {\bf m}_{wl} \Vert^2}
		+ \frac{{\bf d}_w \cdot {\bf A}_w{\bf m}_{wl}}
		{\Vert {\bf d}_w \Vert^2 + \Vert {\bf A}_w {\bf m}_{wl} \Vert^2} \right].
	\end{align}
\end{linenomath*}
\begin{linenomath*}
	\begin{align}\label{eq11}
		\frac{\partial \varPhi}{\partial {\bf m}_{wl}} = &
		- \frac{(\Vert {\bf m}_w \Vert^2 + \Vert {\bf m}_{wl} \Vert^2) \cdot 
			{\bf m}_w^T
			- 2 {\bf m}_w {\bf m}_{wl} \cdot {\bf m}_{wl}} 
		{(\Vert {\bf m}_w \Vert^2 + \Vert {\bf m}_{wl} \Vert^2)^2} 
		\notag
		\\& - \frac{(\Vert {\bf d}_w \Vert^2 + \Vert {\bf A}_w {\bf m}_{wl} \Vert^2) \cdot {\bf A}_w^T {\bf d}_w^T
			- 2 {\bf d}_w ({\bf A}_w {\bf m}_{wl}) \cdot {\bf A}_w^T \cdot ({\bf A}_w {\bf m}_{wl})}
		{(\Vert {\bf d}_w \Vert^2 + \Vert {\bf A}_w {\bf m}_{wl} \Vert^2)^2}.   
	\end{align}
\end{linenomath*}

The gradient of ${\bf m}_{wl}$ with respect to ${\bf m}_l$ can be easily obtained as $\frac{\partial {\bf m}_{wl}}{\partial {\bf m}_l} = {\bf W}_m$. Once the subsurface grid discretization is completed, the model weighting matrix ${\bf W}_m$ is positive definite and no longer changes. At this point, the gradient of $\varPhi$ with respect to ${\bf m}_l$ can be viewed as a scaled version of the gradient of $\varPhi$ with respect to ${\bf m}_{wl}$. Combining Equations \ref{eq11}, we obtain:
\begin{linenomath*}
	\begin{align}\label{eq13}
		\frac{\partial \varPhi}{\partial {\bf m}_l} = 
		\frac{\partial \varPhi}{\partial {\bf m}_{wl}} \cdot \frac{\partial {\bf m}_{wl}}{\partial {\bf m}_l}
		= \frac{\partial \varPhi}{\partial {\bf m}_{wl}} \cdot {\bf W}_m.
	\end{align}
\end{linenomath*}

From Equation \ref{eq13}, the model weighting matrix ${\bf W}_m$ participates in the gradient optimization in a multiplicative manner. This indicates that at each iteration, ${\bf W}_m$ can stably balance the loss terms.

\subsection*{Masked 3-D Modeling of Density Logging Data}
\label{subsec:logging}
\begin{figure}
	\noindent\includegraphics[width=\textwidth]{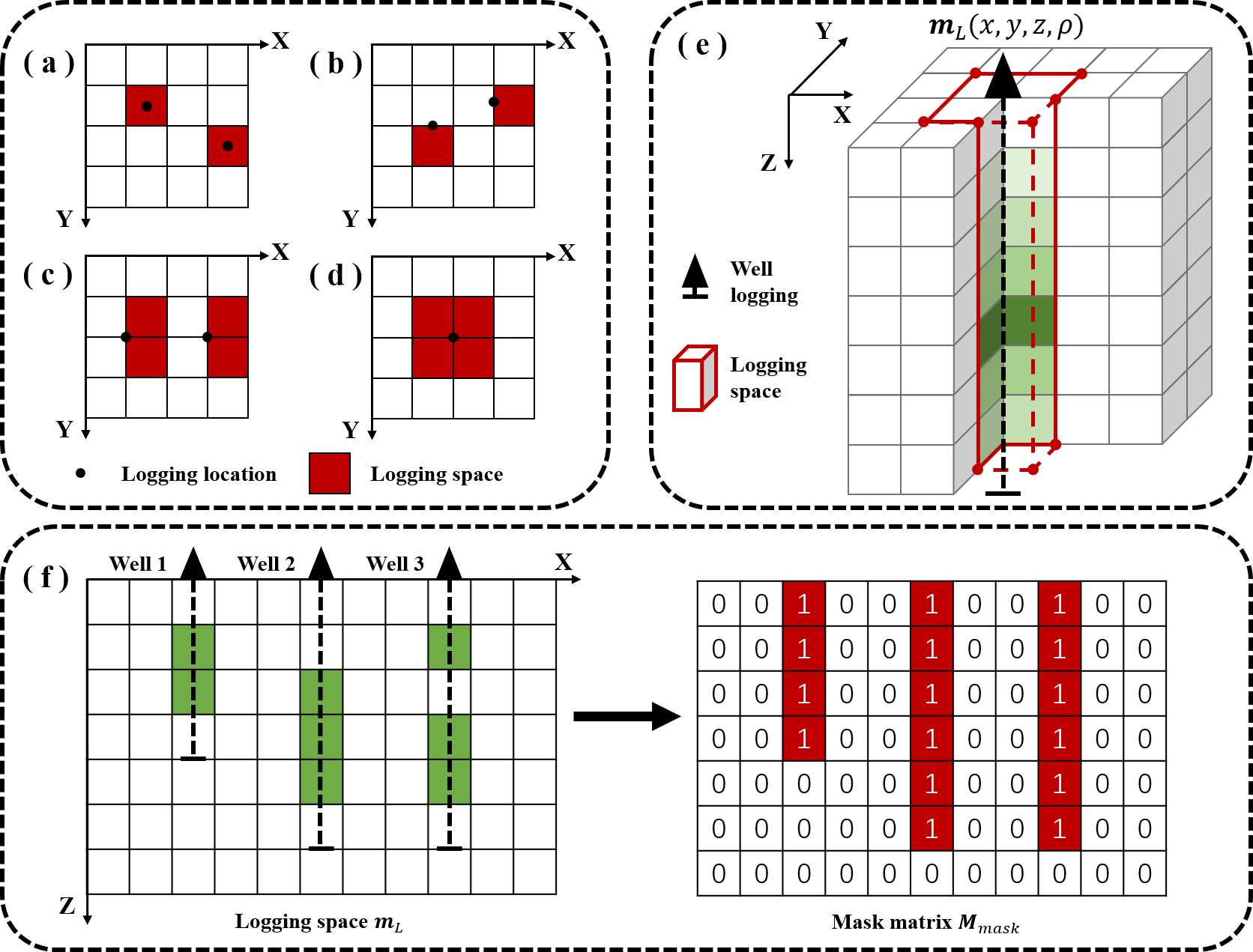}
	\caption{The selection methods for the well logging space: well logging (a) located within the grid, (b) located on the measurement line, (c) densely distributed at the intersection of measurement lines, and (d) sparsely distributed at the intersection of measurement lines. (e) The 3-D well logging space. (f) FG-Mask in well logging space ${\bf m}_L$.}
	\label{fg3}
\end{figure}

The matching of well logging data (centimeter scale) and gravity data (hundred-meter scale) is a manifestation of the multi-source and multi-scale problem in geophysics. Generally speaking, the variation in formation density is continuous, so the mainstream solution is to set the physical property values around the well logging locations to be consistent with the real collected information \cite{vasiljevic2019simple, wu2022discussions, liu2025petrophysical}. This study utilizes the processed formation thickness information derived from well log curves as the well logging data. The set composed of the grids surrounding the well logging is referred to as the well logging space, and its selection criteria for all testing models is illustrated in Figures \ref{fg3}a and \ref{fg3}b, while the selection for field data is demonstrated in Figures \ref{fg3}c and \ref{fg3}d. The specific process of masked modeling is as follows: 

1) Select the well logging space based on the distribution of the well logs. Establish the physical property distribution ${\bf m}_L$ within the well logging space, which will be used as a label during the network training (Figure \ref{fg3}e). Employ a foreground mask approach (FG-Mask), where values within the well logging space are set to 1 and those outside to 0, generating the binary mask matrix ${\bf M}_{\rm mask}$ (Figure \ref{fg3}f).  

2) Apply the masking operation (Equation \ref{eq_mask}) through new mask weighted matrix ${\bf W}_{\rm mask} = {\rm diag}({\bf M}_{\rm mask}) $ to remove the data from the inversion model ${\bf m}_l$ outside the well logging space, retaining only the physical property values ${\bf m'}_L$ recovered within the well logging space.

\begin{linenomath*}
	\begin{align}\label{eq_mask}
		{\bf m'}_L = {\rm diag}({\bf M}_{\rm mask}) {\bf m}_l = {\bf W}_{\rm mask} {\bf m}_l.
	\end{align}
\end{linenomath*}

Finally, the well logging data is incorporated as a constraint term into the NN training by minimizing $\Vert {\bf m}_L - {\bf m'}_L \Vert^2$. During training, this constraint is typically assigned a relatively high weight (see Table \ref{table_hyper-par}).

\subsection*{Dual-Phase Training Approach}
\label{subsec:2phase}
\begin{figure}
	\noindent\includegraphics[width=\textwidth]{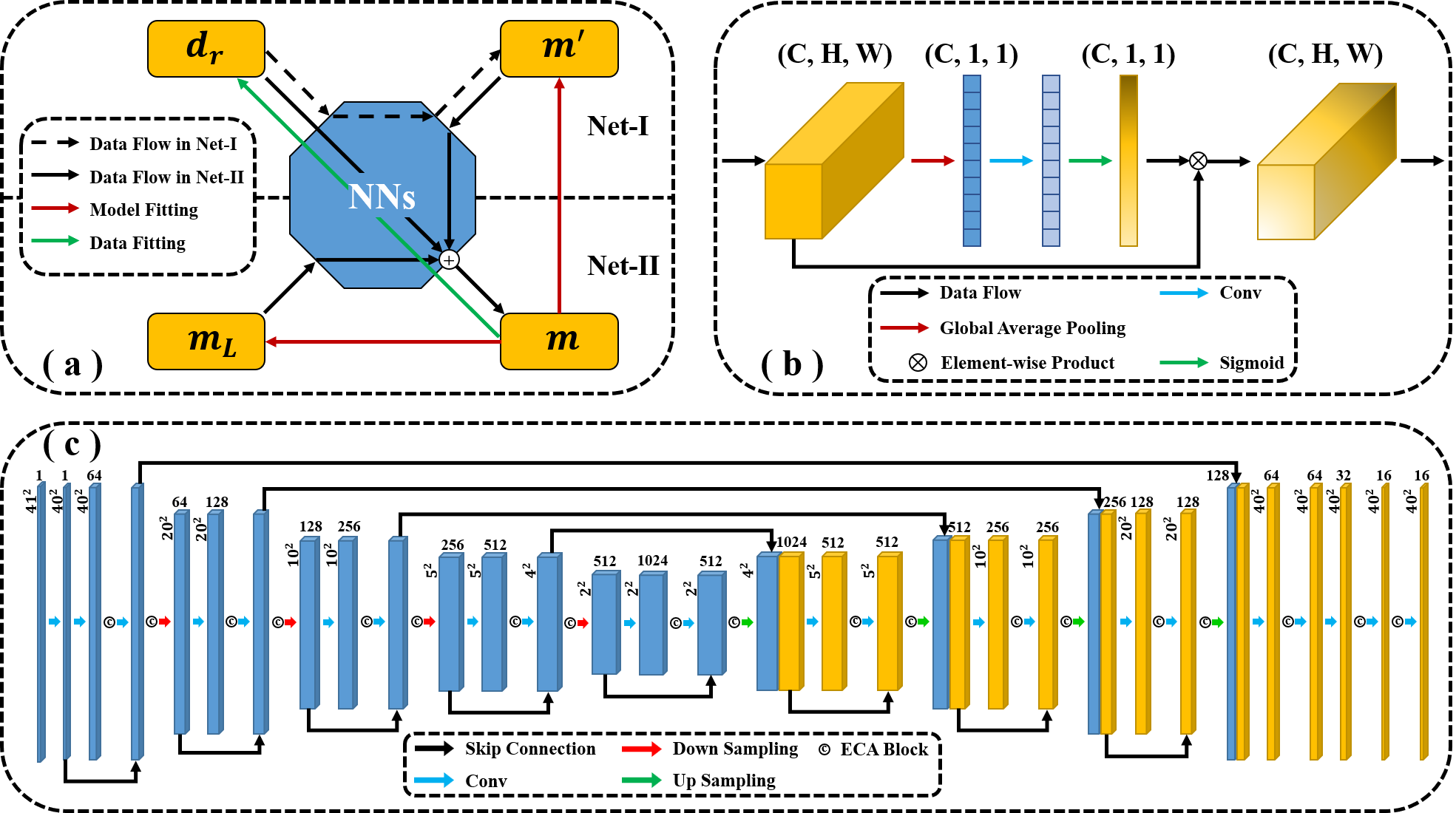}
	\caption{(a) The flowchart of the dual-phase training approach, (b) the ECA block, and (c) the detailed architecture of the ResU-Net network.}
	\label{fg4}
\end{figure}

The flowchart of the dual-phase training approach used in this study is shown in Figure \ref{fg4}a. The detailed steps are as follows:

1) First, use the simulation generated training datasets to train the large pre-trained Net-I, enabling it to learn the mapping from anomaly data to the density model (Equation \ref{eq5}). The loss function for this process is given by Equation \ref{eq10}, where the data constraint term introduces the depth weighting sensitivity matrix ${\bf A}_w$ to the NN. This enables the network to enhance the update magnitude for deeper model parameters while minimizing data misfit, thereby improving depth resolution (detailed discussion in Section Inversion Methodology Based on Depth-weighting). Due to the inherent non-uniqueness of gravity inversion, the output of Net-I can be regarded as the average solution within the training geological datasets space. It provides a reasonable initial model for Net-II's detailed refinement.

2) For the unlabeled field data $\bf {d_r}$, we input it into Net-I to obtain a rough inversion model $\bf {m'}$ (the black dashed arrow in Figure \ref{fg4}a). Then, we use $\bf {m'}$ as a pseudo-label for $\bf {d_r}$ and input them into Net-II, where the inversion result $\bf {m'}$ are further refined under the constraint of prior information (such as the density logging data).

3) We perform 3-D masked modeling of the known density well logging data to obtain both the true density distribution $\bf {m_L}$ and the predicted density distribution $\bf {m'_L}$ within the well logging space (detailed discussion in Section Masked 3-D Modeling of Density Logging Data). 

4) In the training of Net-II, we transfer the network parameters from Net-I and perform full fine-tuning, using $\bf {d_r}$ and $\bf {m'}$ as input sample and label, with $\bf {m_L}$ as the constraint term for enhanced training, to obtain the final output model $\bf {m}$ (the black solid arrow in Figure \ref{fg4}a). During this process, the well logging loss $\Vert {\bf m_L} - {\bf m'_L} \Vert^2$ provides valuable local hard constraints by fixing the inversion results ${\bf m'_L}$ within the well logging space to known measured values ${\bf m_L}$. This serves as a reliable spatial anchor that reduces non-uniqueness by eliminating solutions inconsistent with the known measurements in the well logging space. Simultaneously, it ensures quantitative accuracy at the anchor points, optimizing the average solution of Net-I to align more closely with the true solution. The loss function of Net-II is as follows:

\begin{linenomath*}
	\begin{align}\label{eq14}
		\varPsi ({\bf m}) = 
		\Vert {\bf m}' - {\bf m} \Vert^2
		+ \alpha \Vert {\bf d_r} - {\bf Am} \Vert^2
		+ \beta \Vert {\bf m_L} - {\bf m'_L} \Vert^2
		+ \gamma TV({\bf m}),
	\end{align}
\end{linenomath*}
where ${\rm TV}({\bf m})$ is the 3-D discrete isotropic total variation regularization (TV), which sums the gradient magnitudes across all voxels through differential approximation and effectively preserves edge information within the model to maintain structural continuity. Its expression is as follows:
\begin{linenomath*}
	\begin{align}\label{eq15}
		{\rm TV}({\bf m}) = \sum_{i, j, k}\sqrt{({\bf D}_x {\bf m})_{i, j, k}^2 + ({\bf D}_y {\bf m})_{i, j, k}^2 + ({\bf D}_z {\bf m})_{i, j, k}^2},
	\end{align}
\end{linenomath*}
where ${\bf D}_x$, ${\bf D}_y$, and ${\bf D}_z$ represent the finite-difference operators of model ${m}_{i, j, k}$ along the x-, y-, and z-directions, respectively. Net-II initializes its parameters based on Net-I and flexibly incorporates prior constraints through a composite objective function, enabling refined adjustments to the coarse model output from Net-I. Furthermore, when applied to unseen geological structure, the proposed method yields reasonable inversion model while maintaining stable performance metrics (Section Test-III: Bishop Model), demonstrating generalization capability beyond the training distribution.

\subsection*{Network Architecture Selection}
\label{subsec:net}
This work adopts the ResU-Net as the foundational network architecture. The residual learning mechanism of ResU-Net enables deeper network structures and a more stable training process while preserving the superior detail-capturing ability of U-Net \cite{liu2020common, wang2023enhanced}. More importantly, a notable trend in geophysical data science is the shift from purely data-driven "black box" models towards "gray box" models that incorporate physical laws to enhance interpretability \cite{hu2024three, guo2025deep, pang2025iterative}. In line with this trend, the primary contribution of our study is not to design or select a state-of-the-art inversion network, but to explore a cost-effective pathway for performance enhancement. Specifically, we investigate how to leverage a transfer learning architecture (Net-II) to intelligently incorporate prior knowledge (density logs) for constraining and improving gravity inversion quality. In this context, a classical and widely validated architecture like ResU-Net serves as a clearer demonstrator for the effectiveness and generalizability of our proposed method than an exceedingly complex novel architecture would. Future studies can readily apply our fusion framework to other base networks.

The specific structure of the network is shown in Figure \ref{fg4}c, where each layer consists of a size of $3\times3$ or $2\times2$ convolutional kernel, a batch normalization (BN) layer, and a nonlinear activation function. Additionally, we introduce the efficient channel attention (ECA) block in each layer of the network (Figure \ref{fg4}b), which allows the NN to effectively capture information from cross-channel interactions \cite{wang2020eca}.

\subsection*{Experimental Configuration Details}
\label{subsec:details}
\begin{figure}
	\noindent\includegraphics[width=\textwidth]{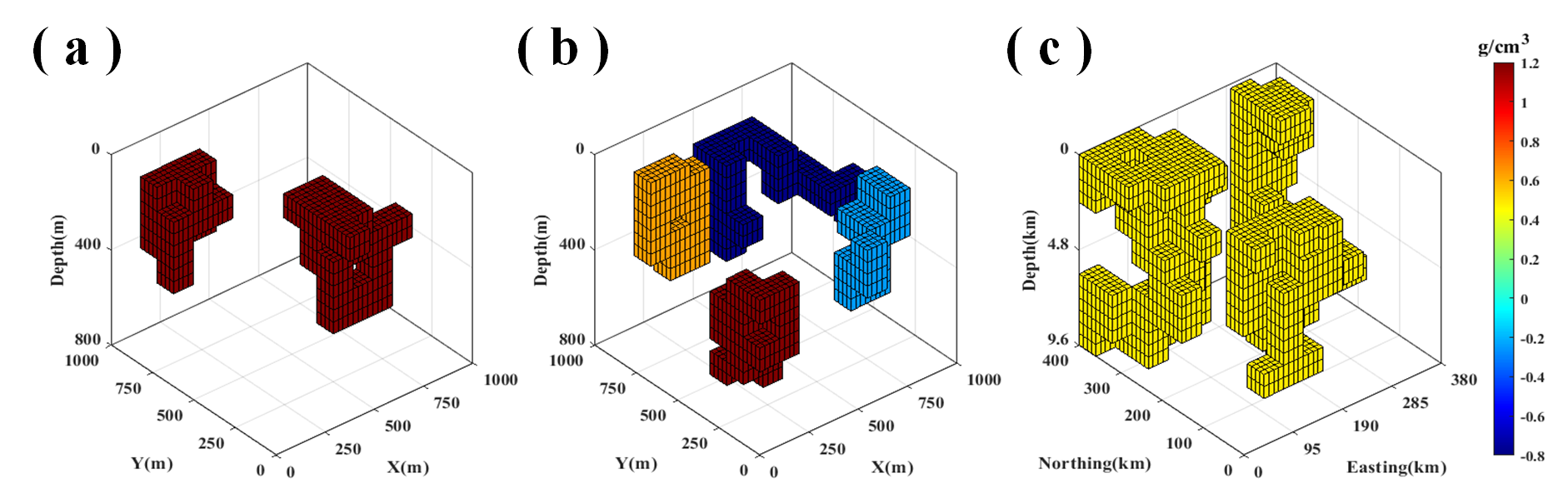}
	\caption{The randomly generated models used in (a) Test-I, (b) Test-II, and (c) Test-III for NN training.}
	\label{train_model}
\end{figure}

We divide the underground space into $40\times40\times16$ cubes ($25 \rm m\times25 \rm m\times50 \rm m$ units each) and generates 30,000 datasets, with a training-to-validation set ratio of 14:1. The datasets are constructed following a strategy similar to \cite{huang2021deep}, whereby multiple random models are generated by controlling the growth iterations of random seeds within the subsurface space. Our method is currently applied only where the training datasets are considered an adequate model of the geology in question. Figure \ref{train_model} displays the randomly generated models used in Test-I, Test-II, and Test-III of the following chapter. Across all tasks in this study, we employ Adam as the optimizer and Table \ref{table_hyper-par} details the hyper-parameters used for specific tasks. The cosine learning rate scheduler is exclusively utilized in Net-I training.

\begin{figure}
	\noindent\includegraphics[width=\textwidth]{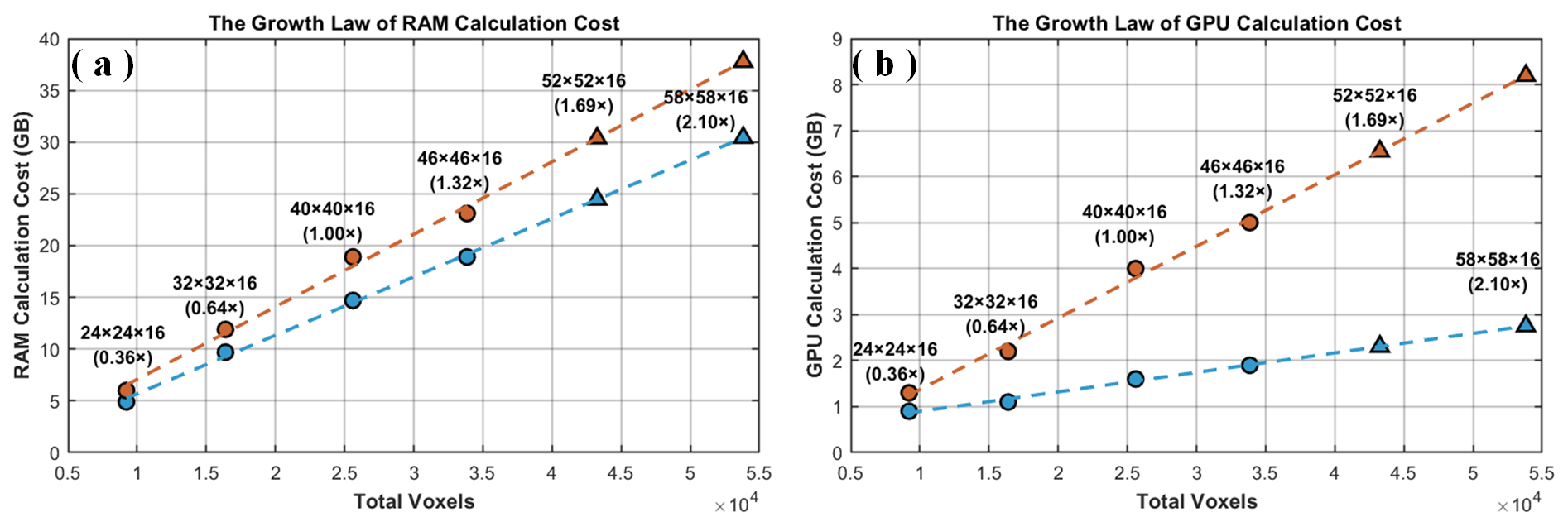}
	\caption{Schematic diagram illustrating the growth laws of (a) GPU and (b) RAM computational costs with increasing model resolution. Blue and orange icons represent the data under the purely data-driven and depth weighting scenarios, respectively. Circular and triangular icons denote the empirically measured data and the theoretically extrapolated data, respectively.}
	\label{voxel}
\end{figure}

The PC configuration includes 1×Intel(R) Core(TM) i7-14700F 2.10 GHz, 1×8 GB NVIDIA GeForce RTX 4060Ti, and 32 GB memory. With this computational capacity, Net-I training consumes 18.9 GB of RAM and 4 GB of GPU. We design four sets of experiments to quantitatively analyze the scaling of computational costs with increasing model resolution under both purely data-driven and depth weighting scenarios. Within the tested sample range, computational costs exhibit a linear relationship with model resolution. Compared to the data-driven case, the introduction of depth weighting increases the GPU cost growth rate from 0.42 GB per 10,000 voxels to 1.6 GB, and the RAM cost growth rate from 5.6 GB to 7 GB per 10,000 voxels. Based on this trend, we reasonably extrapolate theoretical data for two higher-resolution scenarios (Figure \ref{voxel}). This demonstrates that when employing physics constrained methods like depth weighting, even a modest increase in resolution can impose significant computational resource demands, necessitating a more careful balance between accuracy and cost.

\begin{table}
	\caption{Hyper-parameter Configurations for All Tasks}\label{table_hyper-par}
	\setlength{\tabcolsep}{3pt}
	\centering
	\begin{tabular}{c| c| c| c| c| c| c| c}\toprule
		\multicolumn{2}{c|}{\textbf{Task}} & \makecell{\textbf{Learning} \\ \textbf{rate}} & \makecell{\textbf{Weight} \\ \textbf{ratio}} & \makecell{\textbf{Training} \\ \textbf{set size}} & \makecell{\textbf{Early-stopping} \\ \textbf{criteria}} & \makecell{\textbf{Batch}\\\textbf{size}} & \makecell{\textbf{Dropout} \\ \textbf{rate}}\\
		\hline
		\multicolumn{2}{c|}{Net-I} & 3e-4 & 1 $:$ 1 & 28,000 & 1e-4 & \multirow{5}{*}{32} & \multirow{5}{*}{0.50} \\
		\cline{1-6}
		\multirow{4}{*}{Net-II} & Test-I & 5e-3 & 1 $:$ 1 $:$ 10 $:$ 0.04 & \multirow{4}{*}{32} & \multirow{4}{*}{1e-5} &  &  \\
		\cline{2-4}
		& Test-II & 5e-2 & 10 $:$ 2 $:$ 10 $:$ 0.08 &  &  &  &  \\
		\cline{2-4}
		& Test-III & 8e-3 & 8 $:$ 0.10 $:$ 20 $:$ 0.01 &  &  &  &  \\
		\cline{2-4}
		& Real Data & 8e-3 & 1 $:$ 2 $:$ 20 $:$ 0.02 &  &  &  &  \\
		\bottomrule
	\end{tabular}
\end{table}

\section{Testing Model Inversion Results} \label{sec:model_test}
We use three metrics, model accuracy ($\rm{MA}$), data accuracy ($\rm{DA}$), and data error ($\rm{DE}$), to evaluate the quality of the inversion results. Additionally, we employ "MA at $z$" to denote the MA of the inversion model at depth $z$, providing a quantitative measure of depth resolution. The higher the $\rm{MA}$, the closer the inversion model ${\bf m}_l$ is to the true model $\bf m$. The higher the $\rm{DA}$, the more the forward data corresponding to the inversion model ${\bf d}_l$ aligns with the true data $\bf d$. The role of the $\rm{DE}$ is to intuitively reflect the extreme values of the data error. Their expressions are as follows:
\begin{linenomath*}
	\begin{align}\left\{\begin{aligned}\label{eq16}
			\rm{MA} & = \left(1 - \frac{\Vert {\bf m} - {\bf m}_l \Vert^2}
			{\Vert \bf m\Vert^2}\right) \times 100\%, \\
			\rm{DA} & = \left(1 - \frac{\Vert {\bf d} - {\bf d}_l \Vert^2}
			{\Vert \bf d \Vert^2}\right) \times 100\%, \\
			\rm{DE} & = \left({\min}({\bf d} - {\bf d}_l), {\max}({\bf d} - {\bf d}_l)\right).
		\end{aligned}\right.\end{align}
\end{linenomath*}

In synthetic model tests, we refer to the method using the Net-I architecture without data constraints as "ResU", and the method using the Net-I architecture with data constraints (Equation \ref{eq6}) but without depth weighting as "ResU-DC". These serve as the unconstrained data-driven DL baseline and the physics-guided DL baseline for this study, respectively. Both are trained under the same data and computational budget as Net-I. Our method requires building and training a new DL network for each application, with Nets-I and II tuned specifically to an individual case study.

Furthermore, comparative analysis with conventional inversion method can better quantitatively evaluate the performance of our approach. Focusing inversion (FI) is a widely adopted and mature technique in gravity inversion \cite{last1983compact, portniaguine20023, gao2017research}. We select it as the benchmark for conventional regularization methods because of its ability to generate compact, spatially focused sparse models, which not only effectively mitigate non-uniqueness issues but also facilitate clear visual comparisons with DL results.

\subsection{Test-I: Syncline Structure Model} 
\label{subsec:test1}
\begin{figure}
	\noindent\includegraphics[width=\textwidth]{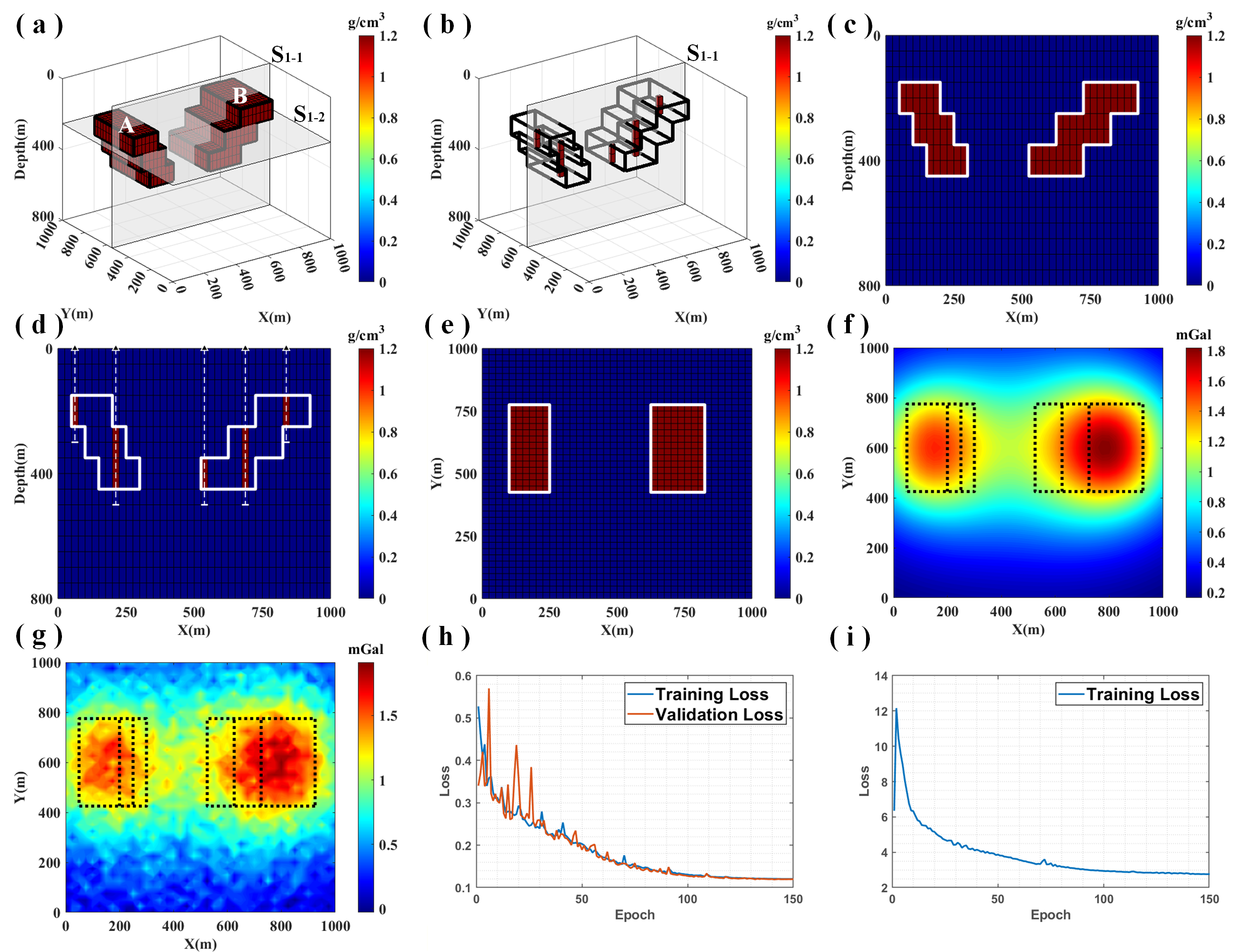}
	\caption{The distribution of (a) the test model and (b) well logging space in 3-D schematic diagram, with the solid lines representing its boundary. The density profiles of (c) the test model and (d) well logging space at positions $\rm S_{1-1}$, where the dashed arrows represent the extension depth of the well logging space. (e) The density profiles of the test model at positions $\rm S_{1-2}$. (f) The gravity anomaly of the test model and its (g) gravity anomaly with 6\% Gaussian noise, with the dashed lines representing its projection on the observation surface. The loss curves of (h) Net-I and (i) Net-II during Test-II training.}
	\label{syncline_true}
\end{figure}

The syncline structure is a common geological model, often associated with mineral resources. We establish a set of syncline structure model to test our method (Figure \ref{syncline_true}a).
The residual density of models A and B is both 1.2$\rm {g/cm}^3$, and the burial depth is 150m. $\rm S_{1-1}$ is a vertical slice, located at Y = 550m. $\rm S_{1-2}$ is a horizontal slice at a depth of 250m. Figure \ref{syncline_true}b shows the selection of the well logging space for the test model and five wells are simulated on the $\rm S_{1-1}$ slice with $\Delta$ = 150m. 

\begin{figure}
	\noindent\includegraphics[width=1\textwidth]{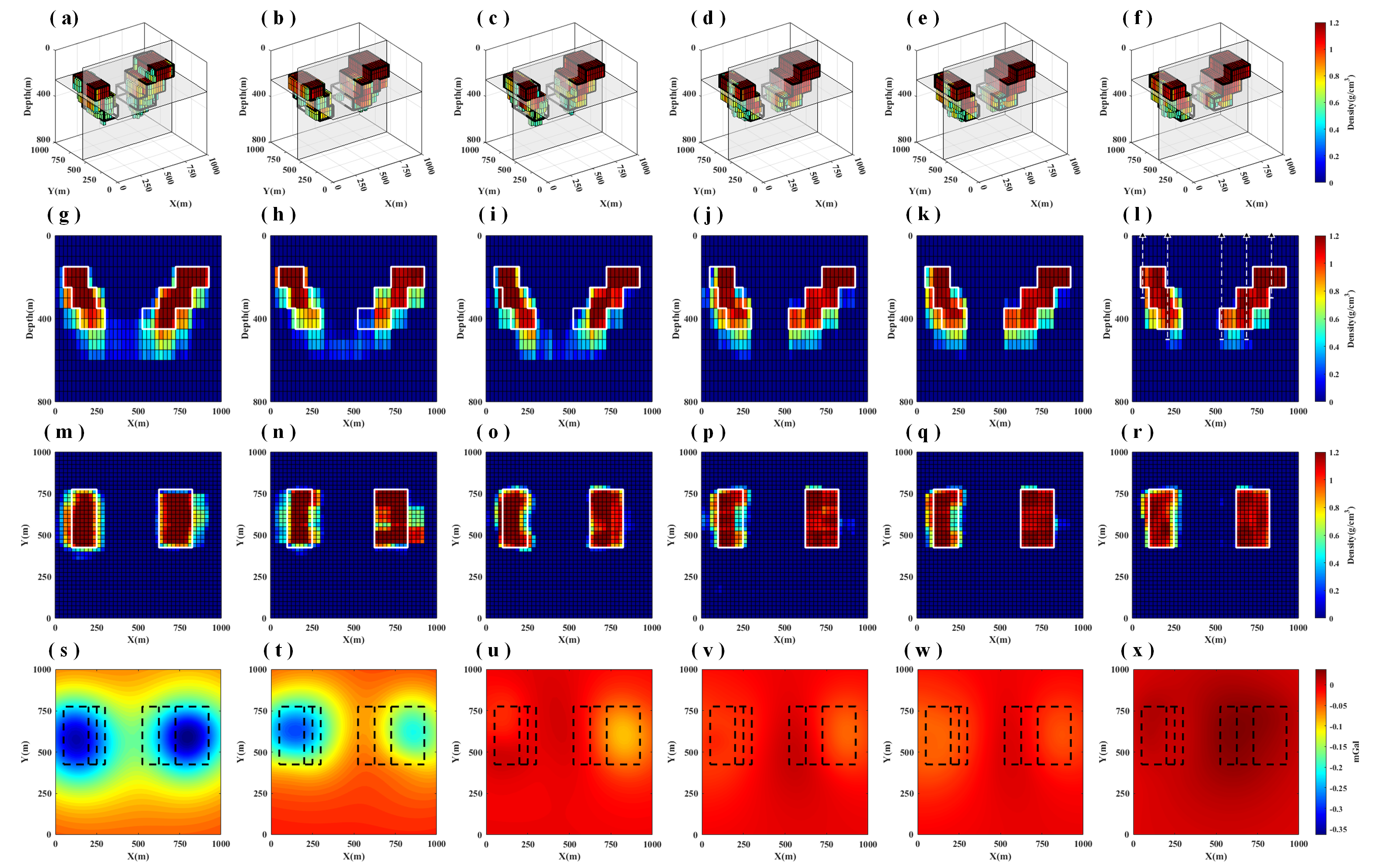}
	\caption{The comparison of inversion results of ablation test. 3-D inversion models from (a) AE0, (b) AE1, (c) AE2, (d) AE3, (e) AE4, and (f) AE5 (showing density > 0.50$\rm {g/cm}^3$). The inversion result slices at $\rm S_{1-1}$ from (g) AE0, (h) AE1, (i) AE2, (j) AE3, (k) AE4, and (l) AE5. Slices at $\rm S_{1-2}$ from (m) AE0, (n) AE1, (o) AE2, (p) AE3, (q) AE4, and (r) AE5. The data fitting error from (s) AE0, (t) AE1, (u) AE2, (v) AE3, (w) AE4, and (x) AE5.}
	\label{syncline_ablation}
\end{figure}

Taking Test-I as an example, we conduct ablation experiments for comparison. The NN is divided into six independent modules (labeled ResU-Net, ECA, Data Constrain, Depth Weighting, TV, and Masked Well Logging) to evaluate the contribution of each module to the inversion. Detailed configurations of the ablation experiments (AE0 to AE5) are summarized in Table \ref{table_AE_setup}, where AE1 corresponds to the ResU method, AE2 to the ResU-DC, AE3 to the Net-I, and AE5 to the Net-II. Table \ref{table_AE_metrics} presents the detailed inversion metrics and evaluation measures of the ablation experiments, with the optimal solutions highlighted in bold. Combined with the inversion results shown in Figure \ref{syncline_ablation}, we analyze and discuss the findings as follows:

AE0 demonstrates the advantages of DL methods in gravity inversion, exhibiting superior depth resolution compared to conventional regularization approaches (higher MA values, with comparative results in Test-IV). Although its DA value is lower, the higher lower bound suggests strong potential for further improvement. The attention mechanism proves to be an efficient way to enhance network performance. In AE1, the introduction of the ECA block increases both parameter count and backward size, yet it also improves inversion quality. Honestly, the network architecture influences the optimization dynamics during inverse problem solving, while the constraint term affects the size and sparsity of the solution space. In AE2, the introduction of data constraints provides physical guidance (the sensitivity matrix ${\bf A}$) to the NN. This effectively enhances the DA value of the inversion results and improves the reconstruction of shallow structures. However, the recovery of deeper features remain relatively blurred. Therefore, in AE3, depth weighting is incorporated to transform the mapping from $\bf{d}$ to $\bf{m}$ into a weighted parameter domain (${\bf {d}}_w$ to ${\bf {m}}_w$), supplying the NN with optimized physical guidance (the weighted sensitivity matrix ${\bf A}_w$). This approach retains the advantages of AE2 while achieving better recovery of deep structures (higher MA values and MA at $z$), albeit at the cost of doubling the training time compared to AE0 and AE1. AE4 initiates the training of the transferred network. Due to the extremely small training set, each training epoch requires only minimal time. By introducing isotropic TV regularization, AE4 not only enhances the physical property values in certain regions but also filters out small anomalous bodies in the inversion results (Figures \ref{syncline_ablation}k and \ref{syncline_ablation}q), which leads to higher MA values. In AE5, prior density logging information is incorporated to constrain the solution space, enforcing consistency between the inverted density distribution and actual well log data within the logging space. As evidenced by Figures \ref{syncline_ablation}j-\ref{syncline_ablation}l, AE5 demonstrates excellent density recovery within the well logging space while positively influencing the physical properties of surrounding anomalies. This approach effectively suppresses spurious anomalies, resulting in optimal MA and DA values among all experiments.

\begin{figure}
	\noindent\includegraphics[width=0.7\textwidth]{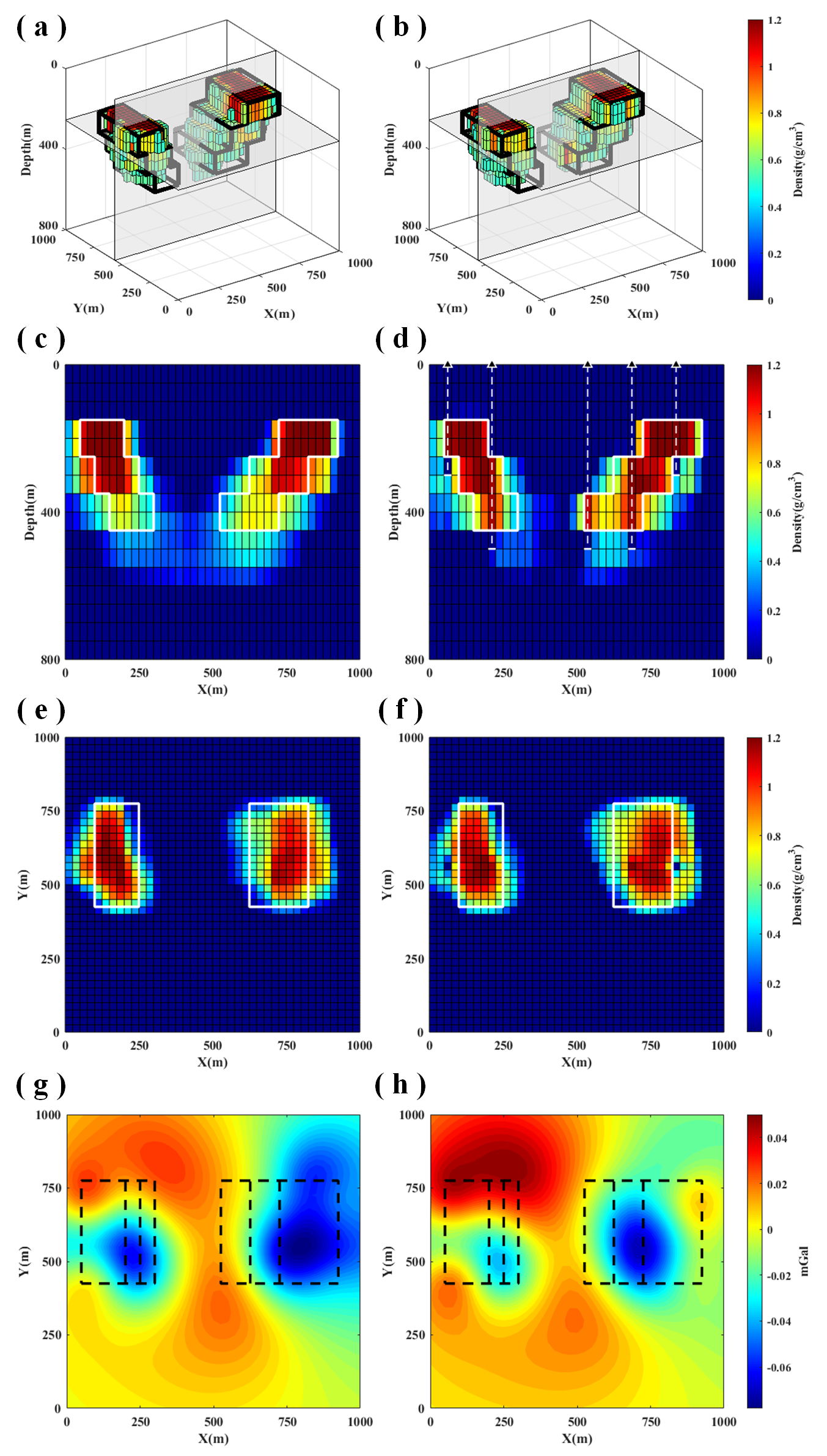}
	\caption{The 3-D inversion models from (a) Net-I and (b) Net-II. The inversion result slices at $\rm S_{1-1}$ from (c) Net-I and (d) Net-II. The inversion result slices at $\rm S_{1-2}$ from (e) Net-I and (f) Net-II. The differences between forward results and true anomalies without noise from (g) Net-I and (h) Net-II.}
	\label{syncline_noise}
\end{figure}

In the processing of real data, effectively suppressing noise is one of the important ways to test the reliability of the method. To simulate a more realistic situation, we add 3$\%$ and 5$\%$ Gaussian noise during network training and add 6$\%$ Gaussian noise to test the noise robustness of our method (Figure \ref{syncline_true}g). It is a normal phenomenon that inversion accuracy decreases as noise increases. Figure \ref{syncline_noise}c demonstrates that Net-I has good noise robustness, with the inversion results clearly showing the contours and trends of the geological bodies. Net-II effectively inherits the noise resistance capability of Net-I, further improving the inversion results under the constraint of 5 wells. It is able to distinguish between the two anomaly bodies, A and B, while also achieving better physical property recovery at greater depths. Table \ref{table_syncline_noise} presents the relevant metrics for noise robustness test. The results demonstrate that Net-II achieves improved inversion quality and enhanced noise robustness with only minimal additional computational cost.

\begin{table}
	\caption{Ablation Experiments Module Setup}\label{table_AE_setup}
	\renewcommand{\arraystretch}{1.01}
	\setlength{\tabcolsep}{5pt}
	\centering
	\begin{tabular}{l c c c c c c c }\toprule
		Method & ResU-Net & ECA & \makecell{Data \\ Constrain} & \makecell{Depth \\ Weighting} & TV & \makecell{Masked \\ Well Logging}\\
		\hline
		AE0 & \ding{51} & \textendash & \textendash & \textendash & \textendash & \textendash \\
		AE1(ResU) & \ding{51} & \ding{51} & \textendash & \textendash & \textendash & \textendash \\
		AE2(ResU-DC) & \ding{51} & \ding{51} & \ding{51} & \textendash & \textendash & \textendash \\
		AE3(Net-I) & \ding{51} & \ding{51} & \ding{51} & \ding{51} & \textendash & \textendash \\
		AE4 & \ding{51} & \ding{51} & \ding{51} & \ding{51} & \ding{51} & \textendash \\
		AE5(Net-II) & \ding{51} & \ding{51} & \ding{51} & \ding{51} & \ding{51} & \ding{51} \\
		\bottomrule
	\end{tabular}
\end{table}

\begin{table}
	\caption{Metrics for Ablation Experiments}\label{table_AE_metrics}
	\renewcommand{\arraystretch}{1.01}
	\setlength{\tabcolsep}{2pt}
	\centering
	\begin{tabular}{l c c c c c c c}\toprule
		Method  & \makecell{Total \\ params} & \makecell{Backward \\ size (MB)} & \makecell{Training time\\(s/epoch)} & MA & \makecell{MA at \\ $z$=250m} & DA & DE(mGal) \\
		\hline
		AE0 & 37,683,927 & 34.42 & 37 & 43.68\% & 55.04\% & 81.13\% & (-0.3642, -0.0393) \\
		AE1(ResU) & 37,684,022 & 40.27 & 50 & 48.60\% & 53.59\% & 88.17\% & (-0.2869, -0.0175) \\
		AE2(ResU-DC) & 37,684,022 & 40.27 & 63 & 52.87\% & 57.39\% & 97.16\% & (-0.0890, 0.0080) \\
		AE3(Net-I) & 37,684,022 & 40.27 & 78 & 60.81\% & 67.08\% & 97.84\% & (-0.0530, 0.0099) \\
		AE4 & 37,684,022 & 40.27 & <1 & 61.29\% & 66.48\% & 97.40\% & (-0.0527, 0.0024) \\
		AE5(Net-II)  & 37,684,022 & 40.27 & <1 & \textbf{64.01\%} & \textbf{69.46}\% & \textbf{98.14\%} & (-0.0012, 0.0376) \\
		\bottomrule
	\end{tabular}
\end{table}

\begin{table}
	\caption{Metrics for Noise Robustness}\label{table_syncline_noise}
	\setlength{\tabcolsep}{10pt}
	\centering
	\begin{tabular}{l c c c c}\toprule
		Method  & MA  & MA at $z$=250m & DA & DE(mGal)\\
		\hline
		Net-I & 38.12\% & 43.38\% & 97.12\% & (-0.0781, 0.0292) \\
		Net-II & \textbf{40.30\%} & \textbf{46.00\%} & \textbf{97.56\%} & (-0.0708, 0.0500) \\
		\bottomrule
	\end{tabular}
\end{table}

\subsection*{Test-II: Multi Density Model} 
\label{subsec:test2}
\begin{figure}
	\noindent\includegraphics[width=0.7\textwidth]{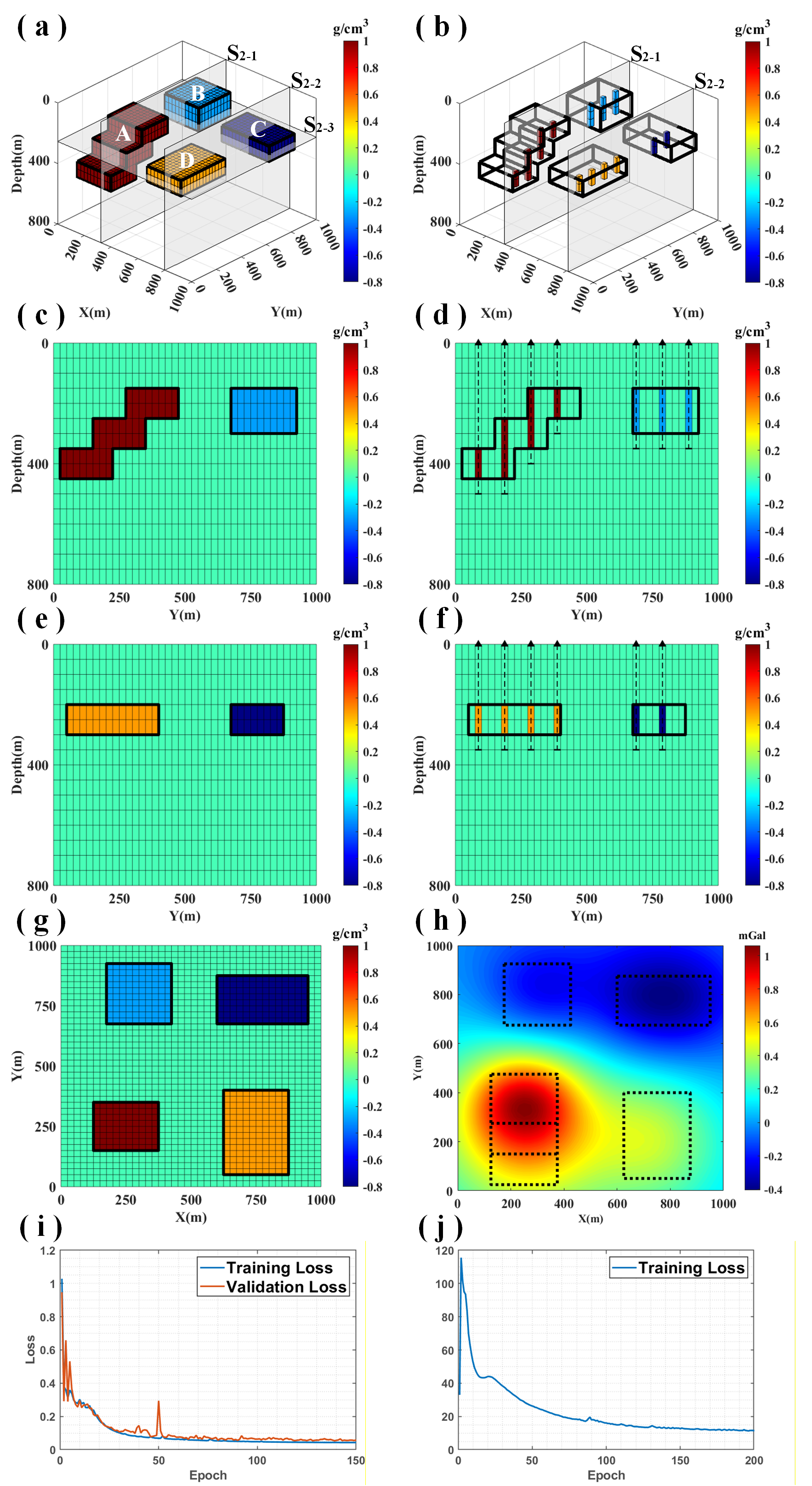}
	\caption{The distribution of (a) the test model and (b) well logging space in 3-D schematic diagram. The density profiles of (c) the test model and (d) well logging space at positions $\rm S_{2-1}$. The density profiles of (e) the test model and (f) well logging space at positions $\rm S_{2-2}$. (g) The density profiles of the test model at positions $\rm S_{2-3}$. (h) The gravity anomaly of the test model. The loss curves of (i) Net-I and (j) Net-II during Test-II training.}
	\label{Md_true}
\end{figure}

Test model consists of four cubes with different sizes and burial depths (Figure \ref{Md_true}a). The residual density of cubes A is 1$\rm {g/cm}^3$, B is -0.3$\rm {g/cm}^3$, C is 0.5$\rm {g/cm}^3$ and D is -0.8$\rm {g/cm}^3$. The burial depth of models A and B is 150m, while the burial depth of models C and D is 200m. $\rm S_{2-1}$ and $\rm S_{2-2}$ are vertical slices, located at X = 325m and X = 800m, respectively. $\rm S_{2-3}$ is a horizontal slice at a depth of 250m. Figure \ref{Md_true}d and \ref{Md_true}f shows the selection of the well logging space for the test model. We simulate 7 wells on the $\rm S_{2-1}$ slice with $\Delta$ = 100m, and 6 wells on the $\rm S_{2-2}$ slice. 

\begin{figure}
	\noindent\includegraphics[width=\textwidth]{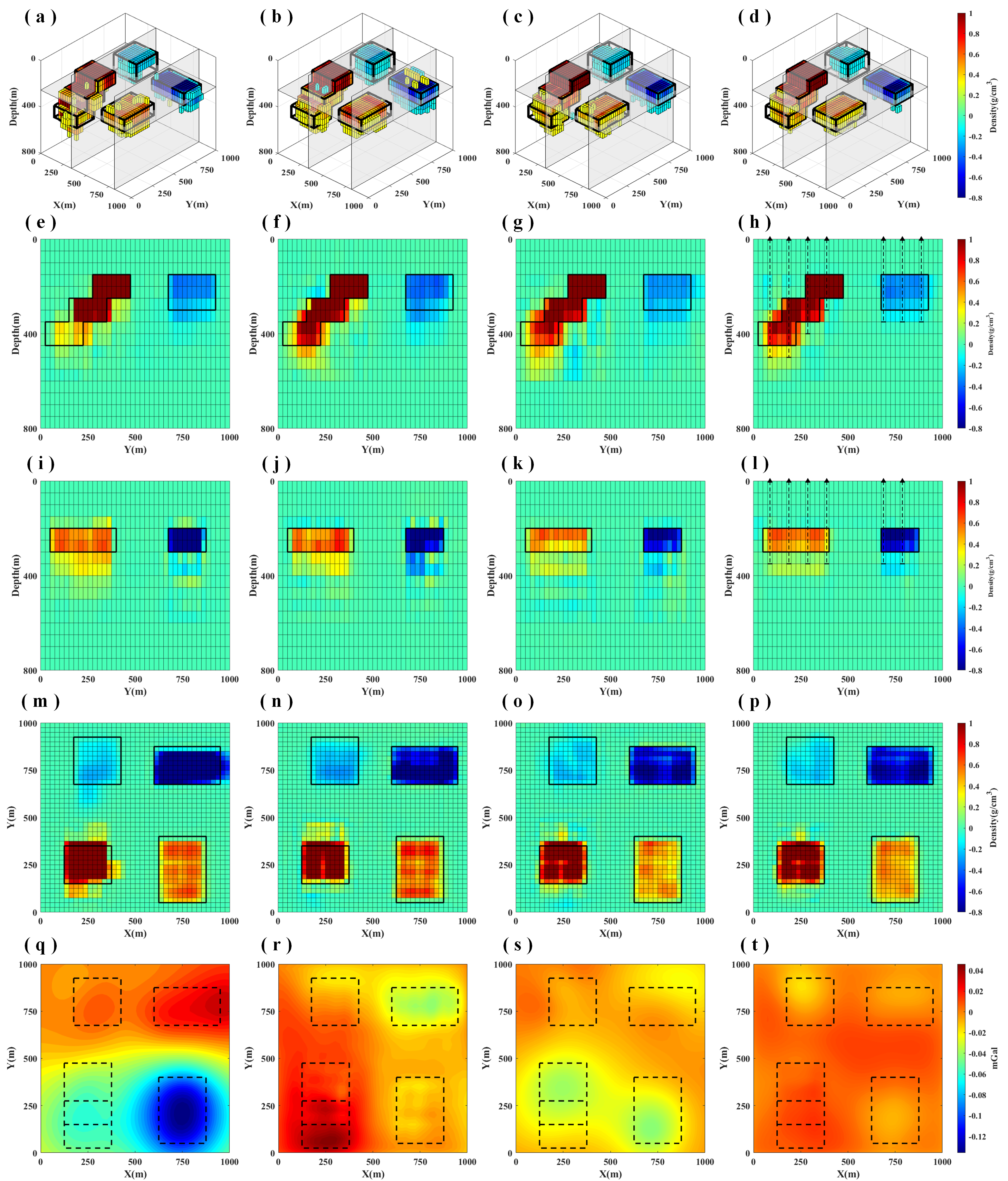}
	\caption{Comparison of inversion results of Test-II. 3-D inversion models from (a) ResU, (b) ResU-DC, (c) Net-I, and (d) Net-II (showing absolute density > 0.20$\rm {g/cm}^3$). Inversion result slices at $\rm S_{2-1}$ from (e) ResU, (f) ResU-DC, (g) Net-I, and (h) Net-II. Slices at $\rm S_{2-2}$ from (i) ResU, (j) ResU-DC, (k) Net-I, and (l) Net-II. Slices at $\rm S_{2-3}$ from (m) ResU, (n) ResU-DC, (o) Net-I, and (p) Net-II. Data fitting error from (q) ResU, (r) ResU-DC, (s) Net-I, and (t) Net-II.}
	\label{Md_comparison}
\end{figure}

As demonstrated in Figure \ref{Md_comparison} and Table \ref{table_md}, compared to the two baseline models, Net-I with depth weighting constraints improves recovery of deep density structures. Subsequently, under the constraint of well logging information, Net-II performs more specific optimization. The inversion model's boundaries are clearer, with fewer false anomalies, and the corresponding data fitting error is lower.

\begin{table}
	\caption{Metrics for Test-II}\label{table_md}
	\setlength{\tabcolsep}{10pt}
	\centering
	\begin{tabular}{l c c c c}\toprule
		Method  & MA & MA at $z$=250m & DA & DE(mGal)\\
		\hline
		ResU & 45.60\% & 54.08\% & 87.44\% & (-0.1361, 0.0365)\\
		ResU-DC & 50.81\% & 59.05\% & 95.87\% & (-0.0415, 0.0462)\\
		Ours(Net-I) & 53.05\% & 62.17\% & 95.50\% & (-0.0433, 0.0056)\\
		Ours(Net-II) & \textbf{55.94\%} & \textbf{62.55\%} & \textbf{98.27\%} & (-0.0172,  0.0156)\\
		\bottomrule
	\end{tabular}
\end{table}

\subsection{Test-III: Bishop Model}
\label{subsec:test3}
\begin{figure}
	\noindent\includegraphics[width=\textwidth]{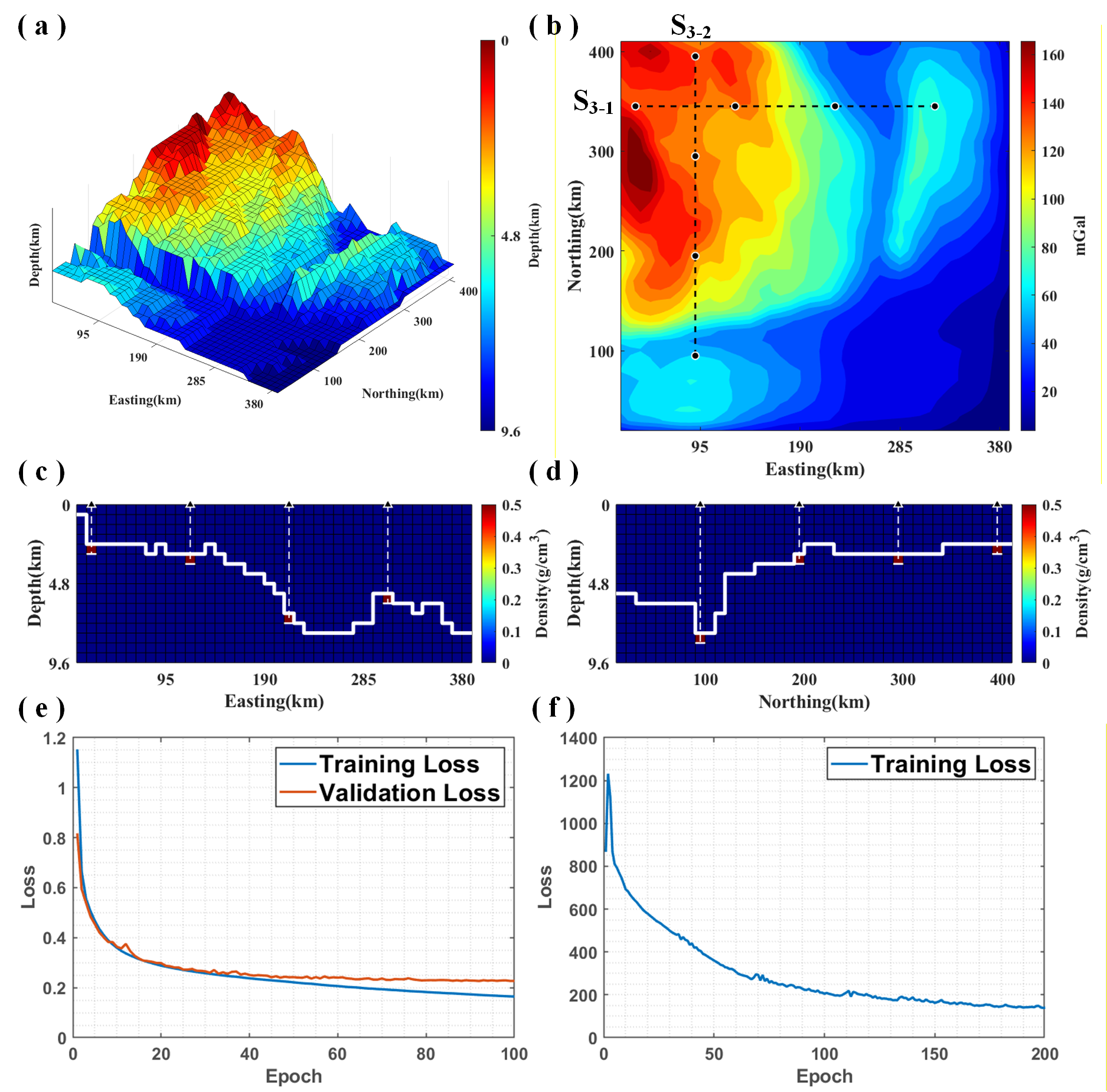}
	\caption{(a)The true basement depth of Bishop Model in 3-D mesh. (b) The corresponding true gravity anomaly of the Bishop Model with black dots indicating assumed well logging positions. The cross-section of the Bishop Model (c) at location $\rm S_{3-1}$ and (d) at location $\rm S_{3-2}$ showing well logging space distributions, where white solid lines denote the true basement top boundary. The loss curves of (e) Net-I and (f) Net-II during Test-III training.}
	\label{Bishop_true}
\end{figure}

To better evaluate the potential of our method for complex practical applications, we apply it to the Bishop Model from the SEG open-source dataset \cite{williams2005comparison}. This model is based on partial topography from the volcanic tablelands area north of Bishop, California, USA, covering an area of 380 km $\times$ 402 km. The depth axis is scaled and shifted to simulate basement morphology at depths ranging from 0.1 km to 9.3 km. The model contains two major offset faults extending along the E-W and N-S directions respectively, forming deep structural lows (Figure \ref{Bishop_true}a). The overall structure exhibits a geological characteristic of shallower northwestern section and deeper southeastern section. The Bishop Model has been widely adopted in basement depth prediction studies \cite{florio2018mapping, jamasb2021multiscale, liu2025integration}.

During data preprocessing, we downsample the Bishop Model to adapt it to our existing NN architecture. The model is discretized into 40 $\times$ 40 $\times$ 16 cubic cells with grid dimensions of 9.5 km $\times$ 10.05 km $\times$ 0.6 km. For the density model configuration, we adopt the similar modeling strategy as \cite{sun2023fast}, setting the residual density to 0 $\rm {g/cm}^3$ for grids above the basement and 0.5 $\rm {g/cm}^3$ below the basement. The corresponding gravity anomaly (Figure \ref{Bishop_true}b) is then computed using fast forward modeling algorithm. Along cross-sections $\rm S_{3-1}$ and $\rm S_{3-2}$, all eight assumed well logs are equally spaced.

\begin{figure}
	\noindent\includegraphics[width=\textwidth]{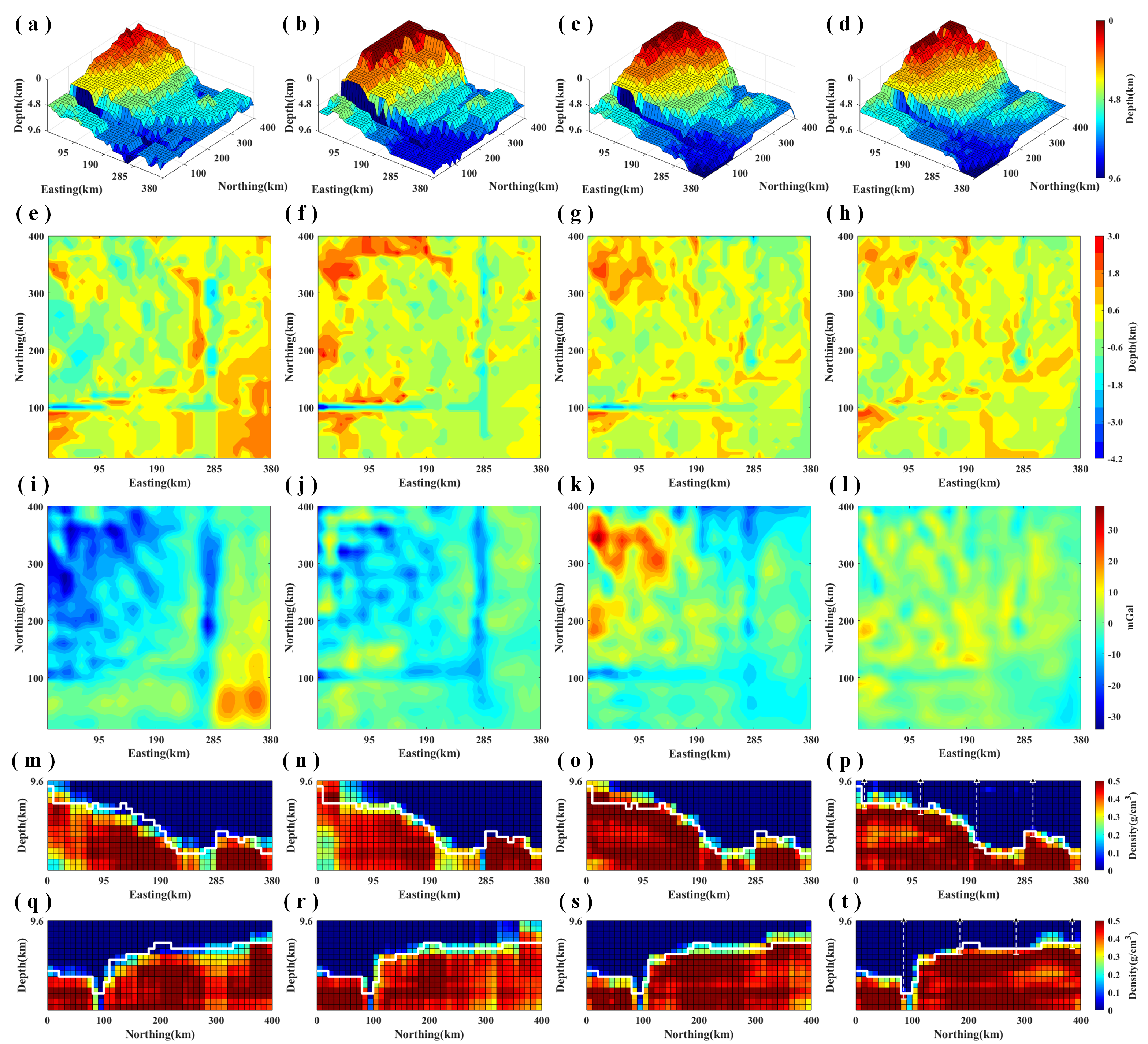}
	\caption{The basement depth prediction results of Bishop Model: 3D mesh views from (a) ResU, (b) ResU-DC, (c) Net-I, and (d) Net-II (showing density > 0.15$\rm {g/cm}^3$), with their residual differences shown in (e-h) against true basement depth and (i-l) against true gravity anomaly. Density inversion cross-sections at location $\rm S_{3-1}$ obtained by (m) ResU, (n) ResU-DC, (o) Net-I, and (p) Net-II. Density inversion cross-sections at location $\rm S_{3-2}$ obtained by (q) ResU, (r) ResU-DC, (s) Net-I, and (t) Net-II.}
	\label{Bishop_comparison}
\end{figure}

In the Bishop Model test, we aim not only for the inversion model to better recover true density values (reflected by higher MA), but also for the corresponding basement depth to match the true depth. Therefore, we propose an additional metric Basement Depth Accuracy (BDA) to evaluate depth prediction performance. Its formulation follows the same structure as MA, but substitutes the true basement depth ${\bf b}$ and predicted basement depth ${\bf b}_l$ for ${\bf m}$ and ${\bf m}_l$ in Equation \ref{eq16}. Table \ref{table_bishop} summarizes the quantitative metrics of all methods applied to the Bishop Model, while Figure \ref{Bishop_comparison} visually demonstrates both the inversion results and basement depth predictions. Figure \ref{Bishop_comparison}a displays the basement prediction results from the ResU method. While generally consistent with the overall geological trend, the predictions exhibit systematically shallower depths in the southeastern section, fail to recover the N-S trending fault, and show significant forward anomaly residuals (reflected in low DA). The density cross-sections further reveal substantial loss of structural details (Figure \ref{Bishop_comparison}m and \ref{Bishop_comparison}q). Figure \ref{Bishop_comparison}b displays the basement prediction results from the ResU-DC method. While the introduction of data constraints improves the DA value of the inversion results, the current form of physical guidance leads to stronger gradients in shallow zones during model updates compared to deeper zones. This imbalance results in poor recovery of deep densities, while excessive sensitivity in shallow layers introduces more spurious anomalies to compensate for data misfit (Figure \ref{Bishop_comparison}n and \ref{Bishop_comparison}r). Consequently, the introduction of physical guidance with more balanced depth weighting is required to mitigate this issue. Figure \ref{Bishop_comparison}c presents the basement prediction results from Net-I, demonstrating better overall alignment with the true topography and preliminary delineation of both major faults. The inversion model exhibits smaller forward anomaly residuals. The density cross-sections in Figure \ref{Bishop_comparison}o and \ref{Bishop_comparison}s reveal that while the inversion results with depth weighting constraints achieve better recovery in deeper zones, some spurious anomalies persist in shallow zones. It leads to an overall shallower predicted basement structure in the northwestern region. This issue could potentially be mitigated by incorporating prior geological information. Figure \ref{Bishop_comparison}d demonstrates Net-II's performance after fine-tuning Net-I's results with constraints from eight synthetic well logs. The inversion model shows superior detail in reconstructing the true basement topography (higher BDA) while generating forward anomaly that better match the true gravity data (higher DA). With well logging constraints, Net-II improves upon Net-I's shallow basement predictions in the northwestern region (Figures \ref{Bishop_comparison}p and \ref{Bishop_comparison}t), effectively suppressing spurious anomalies near the well logging space and producing more focused physical property recovery with sharper boundaries. These results indicate that Net-II achieves significantly enhanced property inversion and basement prediction quality with minimal computational overhead.

Honestly, the Net-II results in this test still exhibit limitations. The predicted basement depths in the northwestern section and along the N-S trending fault are systematically shallower. We attribute these issues primarily to two factors. First, the dual-task challenge of simultaneous density inversion and basement depth prediction exceeds the conventional application scope of the Bishop Model (typically focused solely on depth prediction). Therefore, in density inversion, when the inverted basement density volume fails to fully reach the preset density values, the missing gravity response must be compensated by anomalous response from overlying spurious small-scale anomalies (Figures \ref{Bishop_comparison}o and \ref{Bishop_comparison}s), thereby introducing depth prediction errors. Second, and most critically, the training dataset lacks basement topographic features similar to the Bishop Model, instead employing multiple random models within the subsurface domain to simulate complex geological scenarios (Figure \ref{train_model}c). However, these randomly generated training models lack geologically meaningful contexts, which ultimately limits the network's feature recognition capability for basement structures. Nevertheless, these results indirectly demonstrate our method's strong generalization potential and development prospects.

\begin{table}
	\caption{Metrics for Test-III}\label{table_bishop}
	\setlength{\tabcolsep}{10pt}
	\centering
	\begin{tabular}{l c c c c}\toprule
		Method  & MA  & BDA & DA & DE(mGal)\\
		\hline
		ResU & 64.00\% & 87.84\% & 89.74\% & (-31.7154, 23.6920) \\
		ResU-DC & 65.25\% & 84.84\% & 91.90\% & (-12.9359, 34.2937) \\
		Ours(Net-I) &66.52\% & 88.63\% & 92.37\% & (-21.6696, 37.6756) \\
		Ours(Net-II) & \textbf{69.45\%} & \textbf{91.33\%} & \textbf{95.63\%} & (-13.4864, 16.3304) \\
		\bottomrule
	\end{tabular}
\end{table}

\subsection{Test-IV: Comparison with FI Methods}
\label{subsec:test4}
In this section, we compare the proposed method with FI based on Test-I. Furthermore, to ensure a fair comparison under identical constraints, we incorporate well logging information into the conjugate gradient FI method using the same masked modeling approach, referred to new FI method as well logging constrained focusing inversion (WLC-FI). Detailed specification of WLC-FI method is provided in Appendix A.

In Figure \ref{syncline_FI}, $\rm S_{4-1}$ is a vertical slice, located at Y = 550m. $\rm S_{4-2}$, $\rm S_{4-3}$, and $\rm S_{4-4}$ are horizontal slices, located at a depth of 300m, 350m, and 400m, respectively. $\rm{W_1}-\rm{W_5}$ are five simulated wells.
While the FI produces highly concentrated results, it do not highlight the true dip direction of the truth model well, leading to certain degrees of morphological distortion (Figures \ref{syncline_FI}d, \ref{syncline_FI}j, and \ref{syncline_FI}m). Figure \ref{syncline_FI}e demonstrates that with the exception of $\rm{W_3}$, which is located relatively far from the focused domain, the remaining four wells effectively constrain the density distribution in their respective areas, significantly improving the alignment of the inverted model's dip direction with the true structure. This is particularly evident in Figures \ref{syncline_FI}h, \ref{syncline_FI}k, and \ref{syncline_FI}n, where the well logging constraints produce a clear inward-shifting trend of the inversion results toward the true boundary. With the same well logging constrain, our method can better delineate the boundaries, as demonstrated by the higher MA value (Table \ref{table4}). 

\begin{figure}
	\noindent\includegraphics[width=\textwidth]{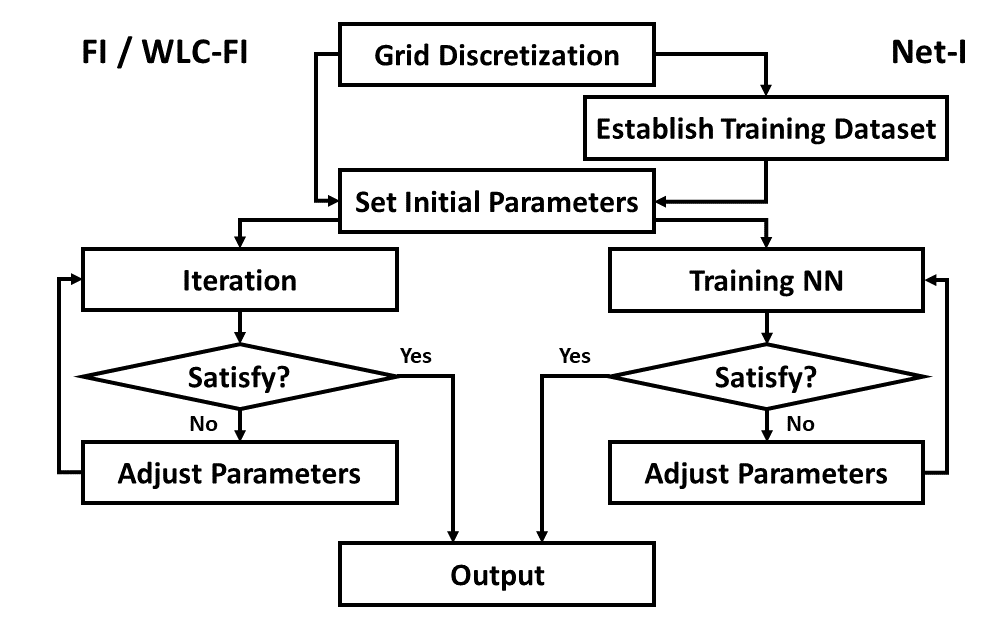}
	\caption{The simplified workflow diagram of Net-I and FI/WLC-FI.}
	\label{workflow_Net1_FI}
\end{figure}

We compare Net-I with FI and WLC-FI using the simplified workflow diagram in Figure \ref{workflow_Net1_FI}. A key distinction lies in the parameter sensitivity and associated tuning costs. The FI and WLC-FI require meticulous manual adjustment of multiple parameters (e.g., focusing factor, regularization weights). This high parameter sensitivity necessitates extensive trial-and-error, often requiring dozens of debugging sessions to achieve a suitable result, making the process tedious and computationally cumulative. In contrast, Net-I exhibits significantly lower parameter sensitivity. While the training of Net-I (2.2 hours for 100 epochs) is more computationally intensive than a single FI iteration (17 min for 100 iterations), the overall cost is balanced by drastically reduced tuning efforts. Key parameters like the learning rate and loss weights can be set to effective values within a few attempts. Furthermore, Net-II trains rapidly (2 minutes for 100 epochs) and maintains this robustness despite having more loss weights. Once trained, the NN provides rapid inversion with high depth resolution for new data. Net-I also addresses the issue of significant data misfit inherent in unconstrained DL methods (higher DA value), and after enhancement through Net-II, the DA value is very close to that of the FI and WLC-FI. In conclusion, all these methods have their own advantages, and the choice depends on the specific goals of the task. If the goal is to obtain a more accurate subsurface density distribution through gravity inversion, and to use it as an initial model for other geophysical methods, then our method would be more suitable.

\begin{figure}
	\noindent\includegraphics[width=0.7\textwidth]{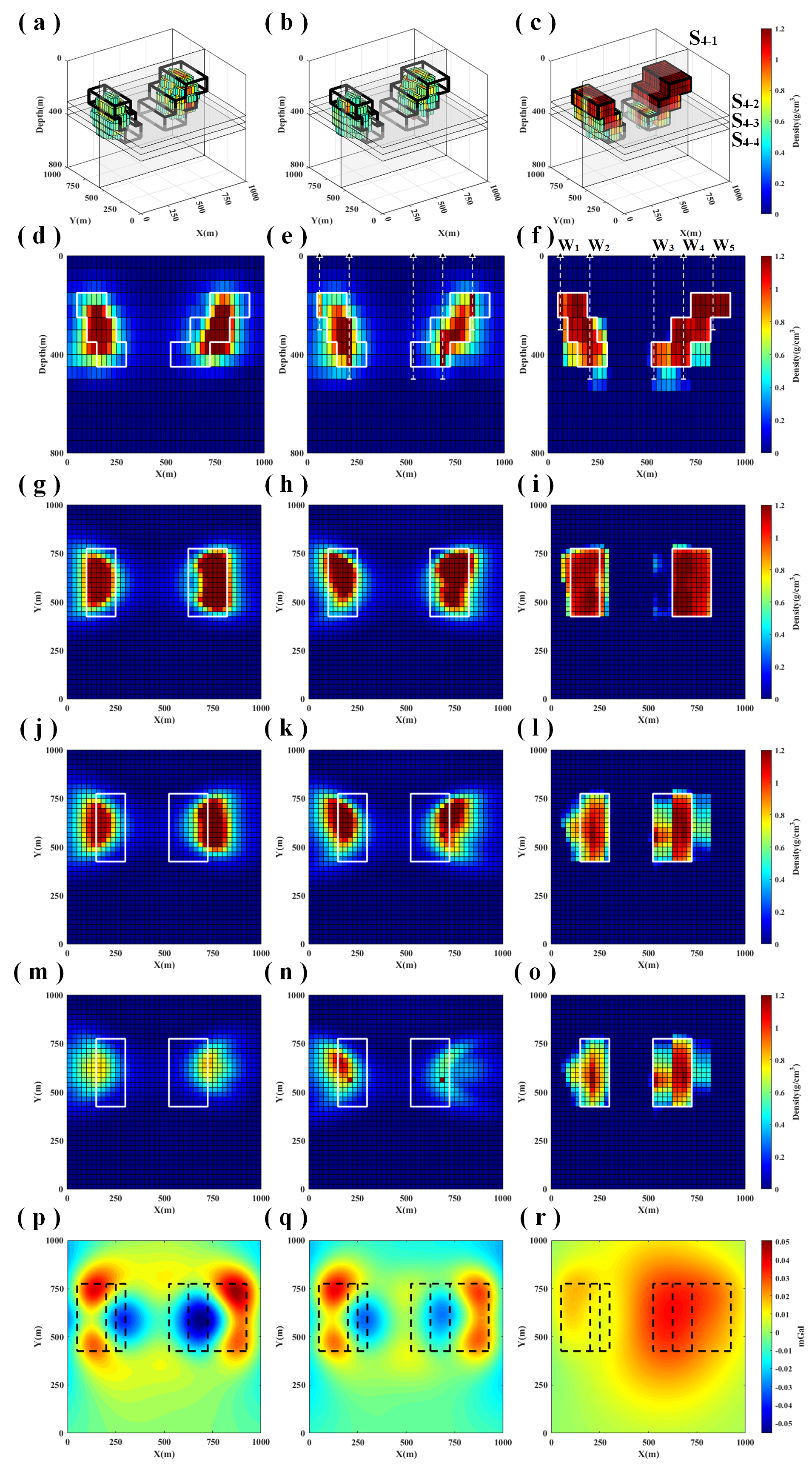}
	\caption{Comparison of inversion results of Test-IV. 3-D inversion models from (a) FI, (b) WLC-FI, and (c) Net-II (showing density > 0.50$\rm {g/cm}^3$). Inversion result slices at $\rm S_{4-1}$ from (d) FI, (e) WLC-FI, and (f) Net-II. Slices at $\rm S_{4-2}$ from (g) FI, (h) WLC-FI, and (i) Net-II. Slices at $\rm S_{4-3}$ from (j) FI, (k) WLC-FI, and (l) Net-II. Slices at $\rm S_{4-4}$ from (m) FI, (n) WLC-FI, and (o) Net-II. Data fitting error from (p) FI, (q) WLC-FI, and (r) Net-II.}
	\label{syncline_FI}
\end{figure}

\begin{table}
	\caption{Metrics for Test-IV}\label{table4}
	\setlength{\tabcolsep}{5pt}
	\centering
	\begin{tabular}{l c c c c c c}\toprule
		Method & MA & \makecell{MA at \\ $z$=300m} & \makecell{MA at \\ $z$=350m} & \makecell{MA at \\ $z$=400m} & DA & DE(mGal)\\
		\hline
		FI & 27.61\% & 44.63\% & 8.39\% & 13.61\% & 98.33\% & (-0.0556, 0.0507) \\
		WLC-FI & 29.04\% & 45.37\% & 13.03\% & 15.77\% & \textbf{98.69}\% & (-0.0322, 0.0406) \\
		Ours(Net-I) & 60.81\% & 67.08\% & 52.82\% & 47.18\% & 97.84\% & (-0.0530, 0.0099) \\
		Ours(Net-II) & \textbf{64.01\%} & \textbf{69.46\%} & \textbf{53.96\%} & \textbf{49.97\%} & 98.14\% & (-0.0012, 0.0376) \\
		\bottomrule
	\end{tabular}
\end{table}

\section{Inversion of Real Data}
\label{sec:real}
\begin{figure}
	\noindent\includegraphics[width=\textwidth]{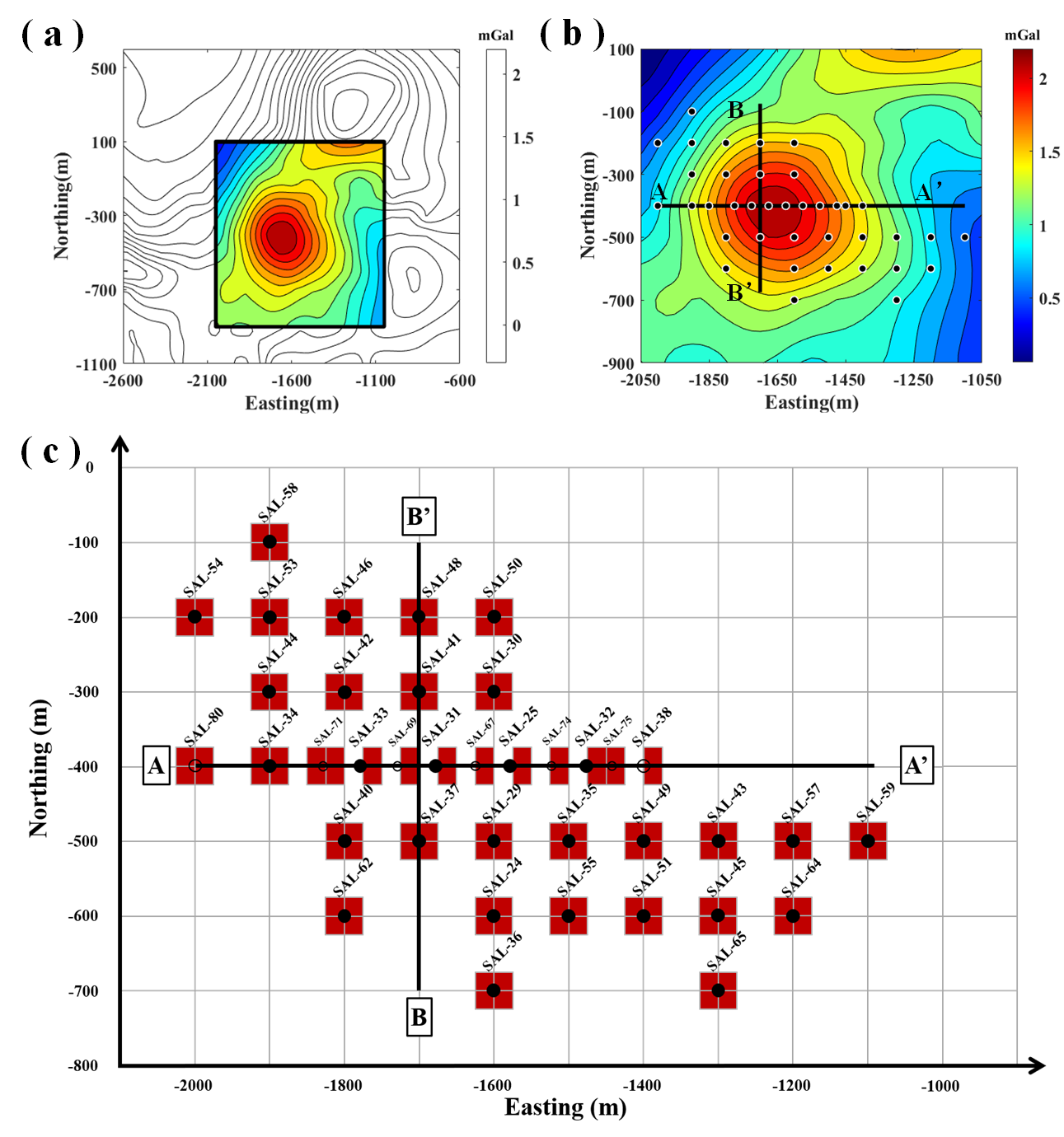}
	\caption{(a) The residual gravity anomaly map of the measured area, with the colored section representing the range selected in this study. (b) The distribution of density well logging (black dots) and two known geological cross-sections AA' and BB' in the study area; (c) The detailed distribution of density well logging locations \cite{johnson1999geology}, where the red area represents the well logging space.}
	\label{fg13}
\end{figure}

\begin{figure}
	\noindent\includegraphics[width=\textwidth]{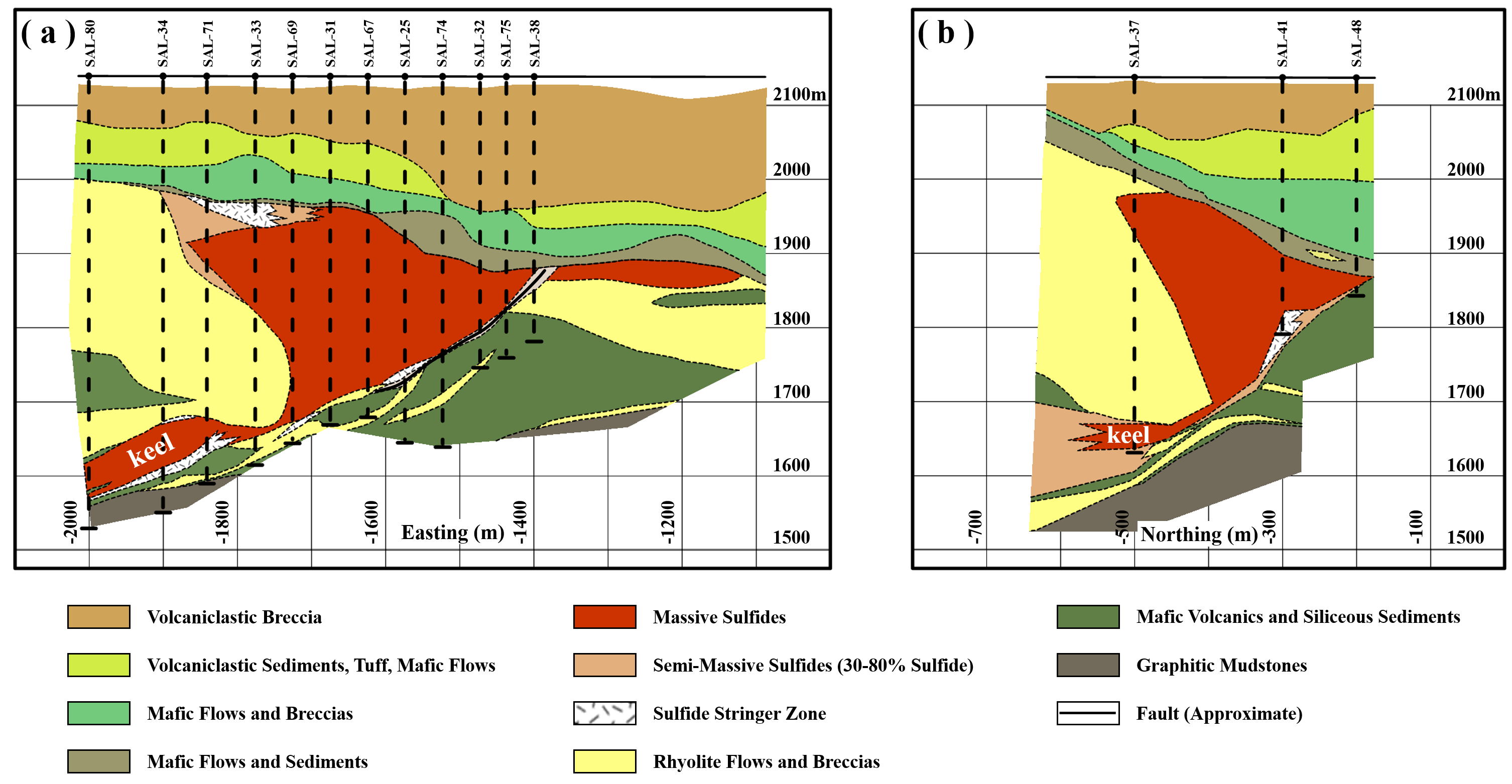}
	\caption{The geological cross-sections at (a) the AA' line, adapted from \cite{johnson1999geology}, and the BB' line which is adapted from \cite{phillips2002geophysical}.}
	\label{fg14}
\end{figure}

San Nicolas sulfide deposit is located in the Zacatecas state of Mexico and its survey area measures approximately 2.0 km $\times$ 1.7 km. \cite{phillips2002geophysical} made significant contributions to the geophysical data acquisition and analysis in this area. To better utilize available geological constraints and improve computational efficiency, we select the colored zone (1.0 km $\times$ 1.0 km) in Figure \ref{fg13}a as our study area. Figure \ref{fg13}b shows the residual gravity anomaly of the study area, where AA' (approximately 900m long, Northing$=$-400m) and BB'(about 500m long, Easting$=$-1700m) represent two known geological cross-sections. According to drilling and core data, the deposit is hosted in iron-magnesium and rhyolitic volcanic rocks. \cite{johnson1999geology} proposed that the main sulfide zone, composed of massive sulfides, is a northwesterly elongate lens, with a thickness of up to 280m, a length of more than 900m, and a width ranging from 200m to over 400m. 
The deposit is almost bounded to the east by a S-W dipping fault, and mineralization develops along the fault, distributed in an unconstrained area of the deposit known as the keel (Figure \ref{fg14}). 
Compared to the surrounding rock (approximately 2.3 $\rm {g/cm}^3$), the massive sulfides have higher density (approximately 3.5 $\rm {g/cm}^3$) and are buried at a depth of about 150m, making it highly suitable for the application of gravity inversion methods \cite{rezaie2017fast, huang2021deep, lv2023fast, wang2024efficient}.

\begin{figure}
	\noindent\includegraphics[width=\textwidth]{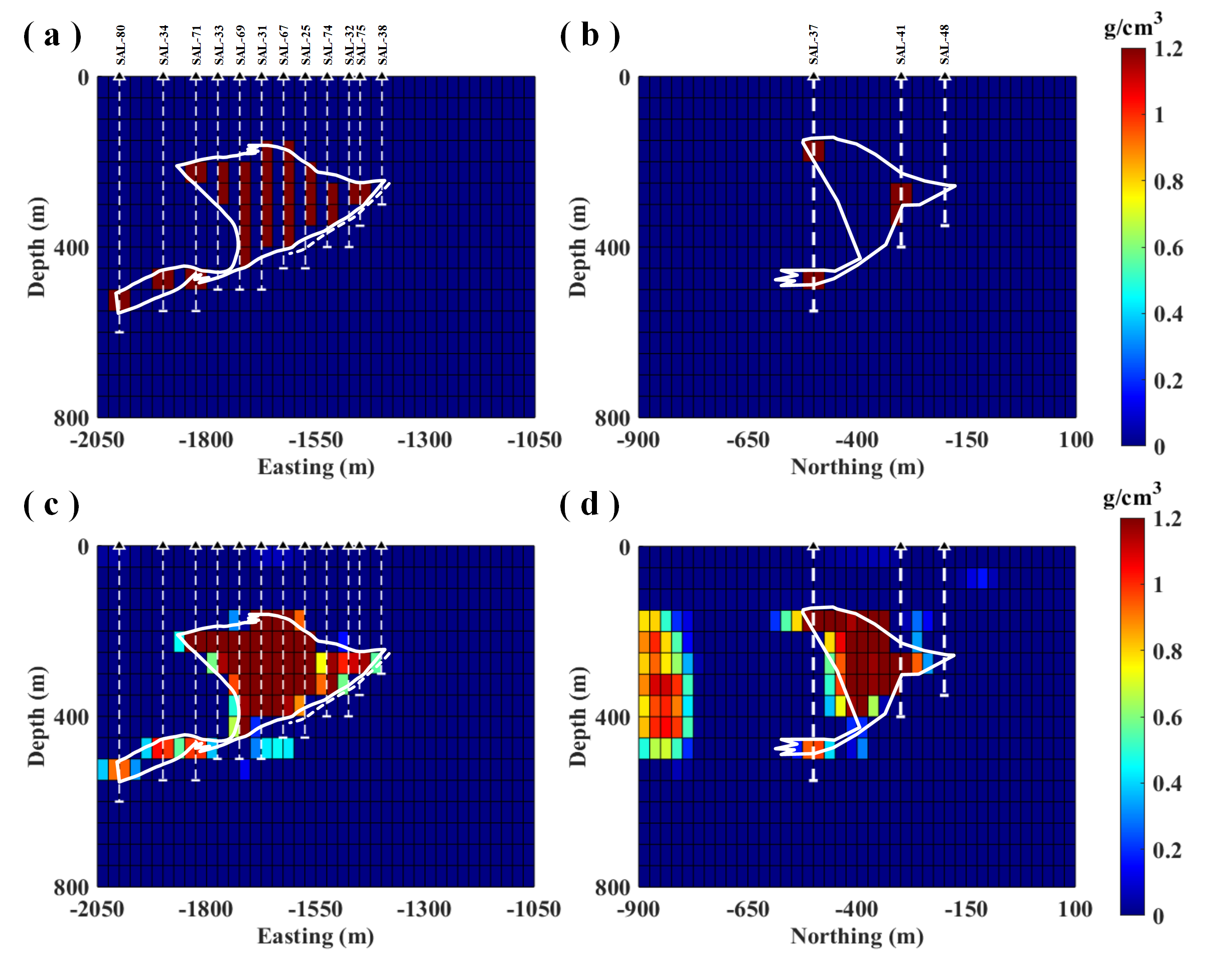}
	\caption{The distribution of the well logging space at (a) AA' and (b) BB'. Inversion result slices at (c) AA' and (d) BB' from Net-II. The white solid line indicates the true outline of the sulfide deposit and white dashed curve indicates the approximate locations of faults.}
	\label{fg15}
\end{figure}

Meanwhile, the San Nicolas area has abundant well logging data (Figure \ref{fg13}c). This study uses information from 38 density wells, with sulfide thickness data collected from 31 wells (solid black dots), as shown in Table \ref{table5}. The remaining 7 wells (hollow black dots) are all located along the cross-section AA', and thickness information can be obtained from Figure \ref{fg14}a. We establish the corresponding well logging space based on the known density well, and its distribution along cross-sections AA' and BB' is shown in Figures \ref{fg15}a and \ref{fg15}b.

Under the constraints of density well logging data, the inversion morphology of the main sulfide zone along cross-sections AA' and BB' shows strong consistency with the actual ore deposit geometry (Figures \ref{fg15}c and \ref{fg15}d), measuring approximately 500 m in E-W length, 250 m in N-S width, and 300 m in thickness. The inversion successfully reconstruct the deposit's eastern sector bounded by S-W dipping faults, aligning with mainstream geological interpretations \cite{phillips2002geophysical}. Moreover, the keel zone, which is a deep mineralized structure along the fault, has been overlooked in most previous studies due to two primary factors: (1) its gravity response is particularly susceptible to being obscured by noise, and (2) conventional discretized grid cells are inadequate for resolving such geometrically intricate features \cite{lelievre2009comprehensive}. However, the four well logs (SAL-34, 37, 71, and 80) provide critical density constraints for the keel zone, enabling us to approximately reconstruct the physical property distribution of the mineralized body within this deep structure.

\begin{figure}
	\noindent\includegraphics[width=\textwidth]{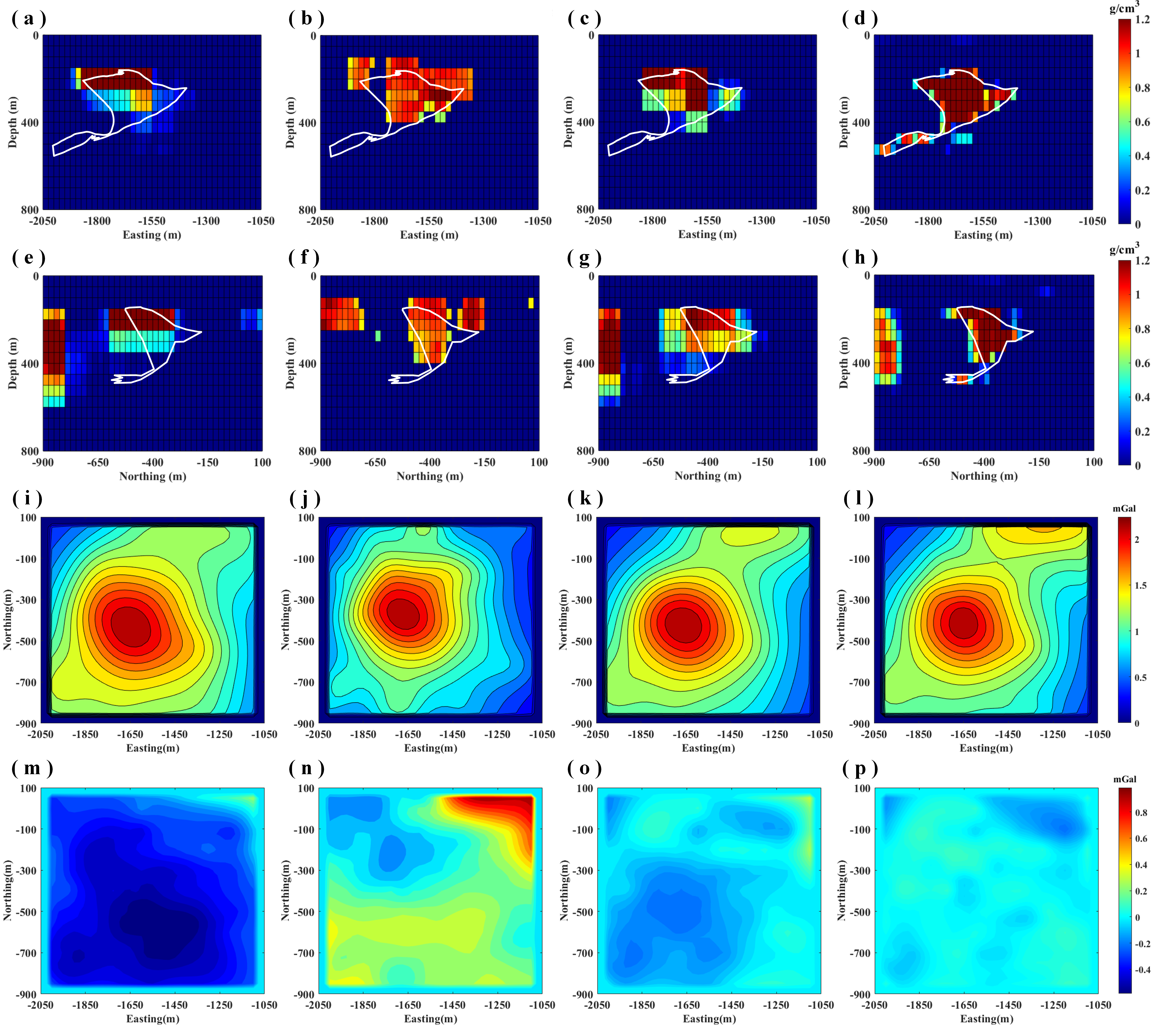}
	\caption{The inversion results obtained by applying different methods to the gravity anomaly data of San Nicolas. The results obtained by \cite{huang2021deep}: (a) slice at AA', (e) slice at BB', (i) forward anomaly, and (m) data fitting error. The results obtained by \cite{lv2023fast}: (b) slice at AA', (f) slice at BB', (j) forward anomaly, and (n) data fitting error. The results obtained by Net-I: (c) slice at AA', (g) slice at BB', (k) forward anomaly, and (o) data fitting error. The results obtained by Net-II: (d) slice at AA', (h) slice at BB', (l) forward anomaly, and (p) data fitting error.}
	\label{SN_comparison}
\end{figure}

To better compare the results, we select two open-source DL methods for testing: Method-I \cite{huang2021deep} and Method-II\cite{lv2023fast}. Both methods are initially developed and applied in the San Nicolas area. We adjust the input size of Method-I from $32\times32\times16$ ($\Delta$ = 50m) to $40\times40\times16$ ($\Delta$ = 25m), while keeping everything else consistent with the original method. The resulting inversion results are also similar to those in the original paper (Figures \ref{SN_comparison}a and \ref{SN_comparison}e). Then, we directly obtain the inversion model of the San Nicolas area from Method-II and extract the research location in this paper for comparison (Figures \ref{SN_comparison}b and \ref{SN_comparison}f). Next, we apply the fast forward modeling method mentioned in this paper to process the inversion models of Method-I and Method-II, obtaining the corresponding forward anomalies. Since the anomaly values at the boundary of the survey area are also related to the geological bodies outside the survey area, to reduce their impact on the test results, we remove the outermost three sets of data from the anomaly map (reducing from $41\times41 = 1681$ to $38\times38 = 1444$).

Table \ref{table6} presents the metrics for this comparison. It can be seen that Method-I and Method-II have relatively low DA values, indicating issues with the inversion results. This is because Method-I has no additional constraints, leading to significant inversion non-uniqueness. Method-II remains an Ud method, it first determines the anomaly shape and location before predicting the density. However, real model is different from test model, making it difficult to accurately determine the subsurface shape. It requires the integration of other prior information to improve performance. As universally recognized, the constraint of prior information can reduce the solution space, thereby minimizing non-uniqueness in the inversion results. Figures \ref{SN_comparison}c and \ref{SN_comparison}g show the inversion results of Net-I in this study. We can see that after introducing the depth-weighting function, the most noticeable improvement is the significant increase in DA values, indicating that the inversion results are more reliable. In comparison, it also has better depth resolution, but still fails to resolve finer structures (such as the keel area within the deposit). This highlights the need for underground information provided by density well logging data to help us achieve more refined inversion results. After introducing density well logging information, we obtain the inversion results of Net-II (Figures \ref{SN_comparison}d and \ref{SN_comparison}h). We successfully recover the density of the keel area, and with good density continuity, the inversion model overall aligns well with the geological body boundaries. Additionally, with higher DA values, indicating that the NN, through the dual-phase training approach proposed in this paper, is capable of producing better and more reliable inversion results.

\begin{table}
	\caption{Sulfide Thickness Information for Wells.}\label{table5}
	\setlength{\tabcolsep}{6pt}
	\centering
	\begin{tabular}{c c c c}\toprule
		\textbf{Well No.} & \textbf{From(m)} & \textbf{To(m)} & \textbf{Thickness(m)}\\ \hline
		SAL-24 & 399.2 & 458.3 & 59.1 \\  \hline
		SAL-25 & 207.1 & 382.5 & 175.4  \\ \hline
		SAL-29 & 188.0 & 496.0 & 308.0  \\  \hline
		SAL-30 & 245.0 & 314.5 & 69.5  \\ \hline
		SAL-31 & 165.7 & 434.0 & 268.3  \\ \hline
		SAL-32 & 247.2 & 329.0 & 81.8  \\  \hline
		SAL-33 & 160.5 & 316.2 & 155.7  \\ \hline
		SAL-34 & 450.6 & 507.5 & 56.9  \\ \hline
		SAL-35 & 223.0 & 415.0 & 192.0  \\ \hline
		SAL-36 & \makecell{405.0} & \makecell{474.5} & \makecell{69.5} \\  \hline
		SAL-37 & \makecell{151.5\\457.7} & \makecell{227.5\\523.4} & \makecell{76.0\\65.7} \\ \hline
		SAL-40 & 471.0 & 510.5 & 39.5 \\ \hline
		SAL-41 & \makecell{232.1\\316.0} & \makecell{354.9\\346.0} & \makecell{122.8\\30.0} \\ \hline
		SAL-42 & 224.7 & 423.0 & 198.3  \\ \hline
		SAL-43 & 211.8 & 335.2 & 123.4  \\ \hline
		SAL-44 & \makecell{216.4} & \makecell{243.6} & \makecell{27.2}  \\ \hline
		SAL-45 & \makecell{297.7} & \makecell{325.7} & \makecell{28.0}  \\ \hline
		SAL-46 & 267.0 & 285.6 & 18.6 \\ \hline
		SAL-48 & 269.0 & 272.4 & 3.4  \\ \hline
		SAL-49 & 226.0 & 378.7 & 152.7  \\ \hline
		SAL-50 & 227.1 & 261.2 & 34.1  \\ \hline
		SAL-51 & 292.4 & 336.3 & 43.9  \\ \hline
		SAL-53 & 277.1 & 335.9 & 58.8  \\ \hline
		SAL-54 & 242.3 & 248.3 & 6.0  \\ \hline
		SAL-55 & \makecell{318.8} & \makecell{346.5} & \makecell{27.7}  \\ \hline
		SAL-57 & 232.0 & 270.1 & 38.1  \\ \hline
		SAL-58 & 265.1 & 270.1 & 5.0 \\ \hline
		SAL-59 & 249.1 & 251.5 & 2.4 \\ \hline
		SAL-62 & \makecell{594.9} & \makecell{628.9} & \makecell{34.0} \\ \hline
		SAL-64 & 180.5 & 226.7 & 46.2\\ \hline
		SAL-65 & 359.0 & 360.9 & 1.9\\ \bottomrule
		\multicolumn{2}{l}{Data come from \cite{johnson1999geology}.}
	\end{tabular}
\end{table}

\begin{table}
	\caption{Metrics for Real Data Inversion}\label{table6}
	\setlength{\tabcolsep}{10pt}
	\centering
	\begin{tabular}{l c c}\toprule
		Method  & DA  & DE(mGal)\\
		\hline
		\cite{huang2021deep}  & 69.72\% & (-0.5846, 0.2545) \\
		\cite{lv2023fast}  & 76.96\% & (-0.1749, 0.9846) \\
		Ours(Net-I)  & 92.03\% & (-0.2136, 0.3180) \\
		Ours(Net-II)  & \textbf{95.72\%} & (-0.2147, 0.1076) \\
		\bottomrule
	\end{tabular}
\end{table}

\section{Discussion}
\label{sec:discussion}
The proposed method allows flexible incorporation of prior constraints (e.g., well logging data or geological structures derived from seismic surveys) during the retraining phase of Net-II, necessitating reliable prior information. As with any geophysical inversion, the well logging data used in this study are subject to various uncertainties, including tool calibration errors, measurement noise, depth registration inaccuracies, and thin-bed upscaling. It must be acknowledged that in practical applications, these uncertainties propagate into the training labels and may, to some extent, reduce the quantitative evaluation scores of the network. Furthermore, to accommodate complex conditions such as well deviation, stratigraphic dip, registration uncertainties, and heterogeneous sampling in field data, adaptive anisotropic masks should be employed based on structural dip and well path, with confidence weighting replacing hard binary masks to reduce the influence of lower-quality intervals. Due to space limitations, the current method treats these data as ideal to establish the theoretical upper limit of performance achievable by introducing such constraints. Thus, a thorough investigation of well logging data uncertainties is warranted in future research.

The quality of Net-I's training significantly determines the inversion of Net-II. However, the randomly generated training models used in the current method lack geologically meaningful scenarios. Therefore, future work should focus on generating tailored training sets that incorporate known geological information from the target area to enhance Net-I's performance. Additionally, the network training involves numerous hyper-parameters, such as loss weighting coefficients ($\alpha, \beta, \gamma$), which are adjusted through iterative experimentation. Although DL method is less sensitive to these parameters compared to conventional regularization methods, significant time is still required for tuning. The rapidly developing field of automatic weight selection methods may substantially reduce the parameter adjustment costs in future work \cite{liu2022auto}.

\section{Conclusion}
\label{sec:con}
Given that current gravity inversion based on the DL generally lack prior constraints, and considering that multi-source physical information integration represents an inevitable trend in development, this study proposes a new method that integrates prior density well logging information to enhance gravity inversion. This method divides the network training into two phases: the pre-training of Net-I and the fine-tuning of Net-II based on it. During Net-I training, depth weighting is introduced to enhance the gradient magnitude for deeper zones during model updates, thereby improving depth resolution and reducing data misfit. In Net-II, density well logging information is incorporated to anchor the inversion results, ensuring consistency with known geological truths. In synthetic model tests, the proposed method demonstrates good depth resolution and noise robustness. When applied to the Bishop Model, our method simultaneously accomplishes both basement depth prediction and density reconstruction tasks, exhibiting stable generalization capability. Compared to the data-driven DL and physics-guided DL baselines, our method achieves average improvements of 10.4$\%$ and 6.8$\%$ in MA, and 8.9$\%$ and 2.4$\%$ in DA, respectively. Subsequently, in the comparison with two FI methods, we evaluate the respective advantages and limitations of both methods, and further discuss the specific scenarios where our method demonstrates superior applicability. Finally, we applied our method to the San Nicolas area in Mexico and compared the results with two DL inversion methods. The results prove that our method effectively restores the shape and physical properties of the anomaly bodies, and is consistent with the known well logging information (DA has been improved by an average of 22.4$\%$).

\section{Data Availability Statement}
\label{sec:available}
All of our data and codes are available at https://zenodo.org/records/17617208.

\section{Appendix A}
\label{sec:appendix}
\subsection{Well Logging Constrained Focusing Inversion}
According to \cite{gao2017research}, the conventional FI method in the depth weighted domain employs the following loss function:

\begin{linenomath*}
	\begin{align}\label{FI}
		\varPsi ({\bf m}_w)_{\rm FI} = 
		\Vert {\bf A}_w {\bf m}_w - {\bf d}_w \Vert^2
		+ \lambda \Vert {\bf m}_w \Vert^2.
	\end{align}
\end{linenomath*}

Note that here ${\bf A}_w = {\bf W}_d {\bf A} {\bf W}_m^{-1} {\bf W}_e^{-1}$, ${\bf m}_w = {\bf W}_e {\bf W}_m {\bf m}$. The ${\bf W}_e$ is the core operator in FI, which is a diagonal matrix with elements of $w_e(m)=(m^2+e^2)^{-1/2}$ along its main diagonal. $e$ denotes the focusing factor, $\lambda $ is damping coefficient. The loss function for WLC-FI is defined as follows:

\begin{linenomath*}
	\begin{align}\label{WLC-FI}
		\varPsi ({\bf m}_w)_{\rm WLC-FI} = 
		\Vert {\bf A}_w {\bf m}_w - {\bf d}_w \Vert^2
		+ \lambda \Vert {\bf m}_w \Vert^2
		+ \mu \Vert {\bf W}_{\rm mask} {\bf m}_w - {\bf W}_e {\bf W}_m {\bf m}_L \Vert^2,
	\end{align}
\end{linenomath*}
where $\mu $ is another damping coefficient that decays exponentially as $e^{-2n}$ with $n$ denoting the iteration number. Through extensive testing, the optimal values are determined as $n$ = 73, $e$ = 0.06, $\lambda $ = 0.006 and $\mu$ = 2.5. The gradient of $\varPsi ({\bf m}_w)_{\rm WLC-FI}$ with respect to the weighted density ${\bf m}_w$ is given by:

\begin{linenomath*}
	\begin{align}\label{gradient-WLC-FI}
		\frac{\partial \varPsi ({\bf m}_w)_{\rm WLC-FI}}{\partial {\bf m}_w} =  ({\bf A}_w^T {\bf A}_w + \lambda {\bf I} +\mu {\bf W}_{\rm mask}^T{\bf W}_{\rm mask}){\bf m}_w -({\bf A}_w^T {\bf d}_w + \mu {\bf W}_{\rm mask}^T {\bf W}_e {\bf W}_m {\bf m}_L).
	\end{align}
\end{linenomath*}

\bibliographystyle{unsrt}
\bibliography{Reference}

\end{document}